\documentclass[iop]{emulateapj}

\newcommand{\kms}{km s$^{-1}$}
\newcommand{\msun}{$M_{\sun}$}
\newcommand{\mJybeam}{mJy\thinspace beam$^{-1}$}
\newcommand{\atc}{atoms\thinspace cm$^{-2}$}
\newcommand{\sixcm}{$\lambda 6$ cm}
\newcommand{\HI}{\ion{H}{1}}
\newcommand{\Halpha}{H$\alpha$}

\newcommand{\sigmav}{$\sigma_{\rm v}$}
\newcommand{\Jybeamms}{Jy\thinspace beam$^{-1}$ m\thinspace s$^{-1}$}

\shorttitle{Observations of NGC 3145}
\shortauthors{Kaufman et al.}

\begin{document}

\title{A Warp in Progress: \ion{H}{1} and Radio Continuum Observations of  
the Spiral NGC 3145}


\author{Michele Kaufman\altaffilmark{1},
Elias Brinks\altaffilmark{2},
Curtis Struck\altaffilmark{3},
Bruce G. Elmegreen\altaffilmark{4},\\
and Debra M. Elmegreen\altaffilmark{5}
}


\altaffiltext{1}{110 Westchester Rd, Newton, MA 02458; email:\\ 
 kaufmanrallis@icloud.com}

\altaffiltext{2}{University of Hertfordshire, Centre for Astrophysics Research,
College Lane, Hatfield AL10 9AB, UK; email: E.Brinks@herts.ac.uk}

\altaffiltext{3}{Department of Physics and Astronomy, Iowa State University,
Ames, IA 50011; email: curt@iastate.edu}

\altaffiltext{4}{IBM Research Division, T.J. Watson Research Center, 
1101 Kitchawan Rd., Yorktown Heights, NY 10598; email: bge@us.ibm.com}

\altaffiltext{5}{Department of Physics and Astronomy, Vassar College, 
124 Raymond Av., Poughkeepsie, NY 12604; email: elmegreen@vassar.edu}

\begin{abstract}
VLA \HI\ observations and \sixcm\ radio continuum observations
are presented of  the barred-spiral galaxy NGC 3145.
 In optical images NGC 3145 has stellar arms that appear to cross,
 forming ``X''-features.
Our radio continuum observations rule out shock fronts at three of
the four ``X'' features, and our \HI\ data  provide 
evidence of gas motions perpendicular to the disk of NGC 3145.
 In large portions of 
NGC 3145, particularly in the middle-to-outer disk,
the \HI\ line-profiles are skewed. Relative to the disk,
the gas in the skewed wing of the line-profiles has $z$-motions 
away from us on the approaching side of the galaxy and 
$z$-motions of about the same magnitude ($\sim 40$ \kms) 
toward us on the receding side.
These warping motions 
 imply that there has been a  perturbation
with a sizeable component perpendicular to the disk over large spatial scales.
Two features in NGC 3145 have velocities indicating that they are
out-of-plane tidal arms. One is an apparent branch of a main spiral
arm on the northeastern side of NGC 3145;  the velocity of the 
branch is $\sim 150$ \kms\ greater than the spiral arm where they appear to
intersect in projection. The other is the arm on the southwestern 
side that forms three of the ``X''-features. It differs in velocity by
$\sim 56$ \kms\ from that of the disk at the same
projected location.
\HI\ observations are presented also of the two small companions NGC 3143 
and PGC 029578.  Based on its properties (enhanced SFR, \HI\ emission
50\% more extended on its northeastern side, etc.), 
NGC 3143 is the more likely of the two companions to have
 interacted with NGC 3145 recently. A simple analytic model demonstrates 
that an encounter between NGC 3143 and NGC 3145 is a plausible
explanation for the observed warping motions in NGC 3145.
\end{abstract}




\section{Introduction}

\object{NGC 3145} at a distance of 54.8 Mpc is a barred spiral galaxy
 with grand-design spiral arms and some peculiar stellar morphology. 
In the Hubble Atlas of Galaxies, \citet{sandage61} writes about NGC 3145 
``There is a 
single faint arm in the southwest quadrant which crosses one of the regular
 arms nearly at right angles. This is a very rare feature of galaxies ...''
We were intrigued by this description and by the presence of three
additional places where stellar arms also appear to
cross, forming ``X''-features, in NGC 3145. The goal of this paper is to understand
these puzzling features by investigating what is happening 
in the interstellar gas at the ``X''-features and elsewhere in the galaxy.

NGC 3145 has two smaller companions:  the barred spiral
\object{NGC 3143} and  the Sdm galaxy \object{PGC 029578}. We refer
to these three galaxies as the NGC 3145 triplet.

\begin{figure*}
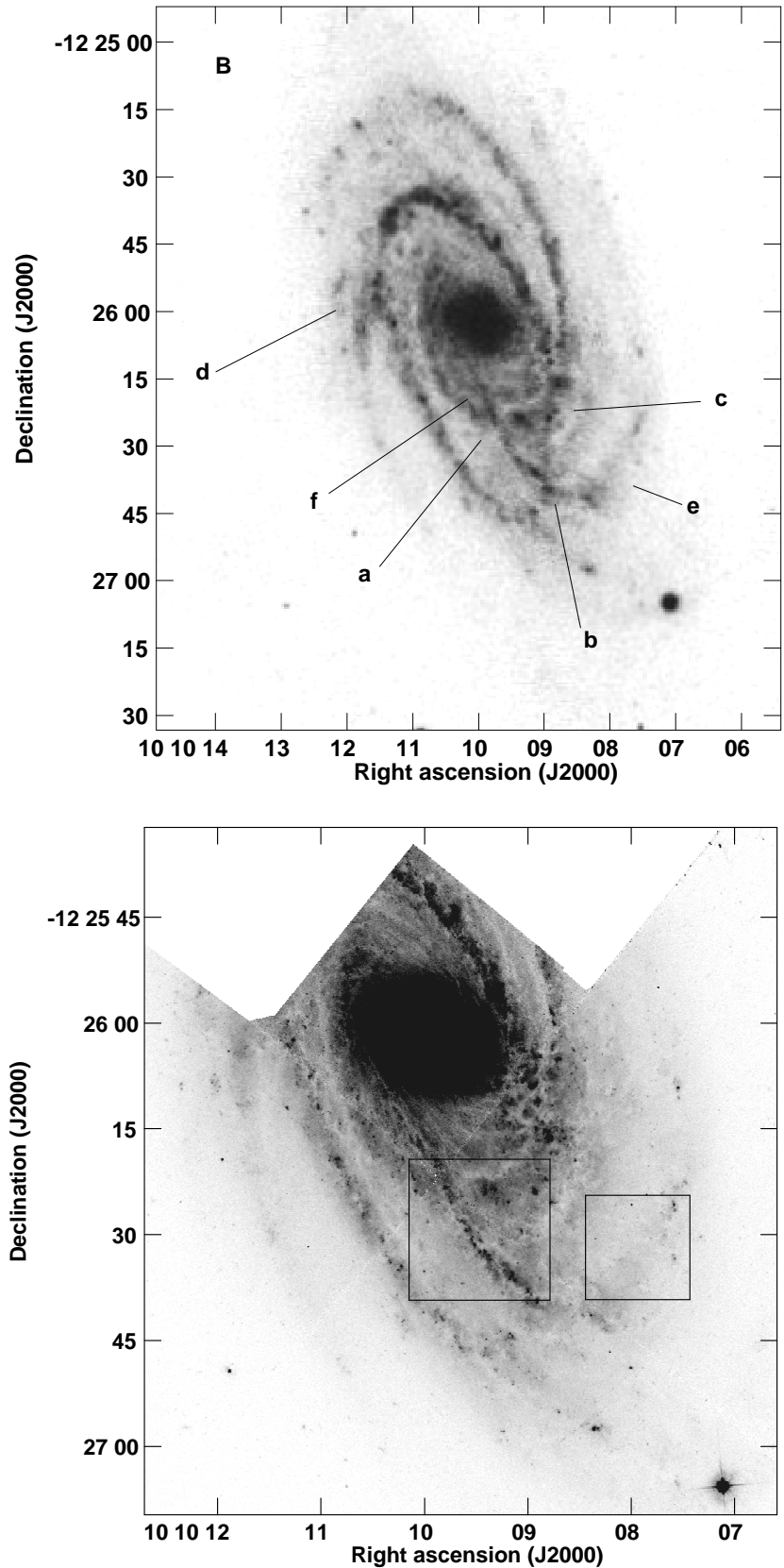

\epsscale{0.75}
\plotone{fig1top.eps}
\plotone{fig1bot.eps}
\caption{Top: $B$ image of NGC 3145 with some optical anomalies 
marked, e.g. ``X''-features
where arms appear to cross at apices $a$, $b$, and $c$
 of the {\it triangle} and at location $f$. Sandage's
peculiar arm outlines the eastern edge of the {\it triangle} from
apex $a$ to apex $b$. The apparent {\it branch} of a main spiral arm 
is labelled $d$, and the western antenna is labelled $e$.
Bottom: {\it HST} WFPC2 $R$ image of NGC 3145. \HI\ line-profiles
for the regions enclosed by the boxes will be displayed and discussed
in Section 4.
\label{B.Features}}
\end{figure*}

The top panel in Figure\,\ref{B.Features} displays a $B$-band image of NGC 3145 
with some of the optical oddities marked,  and the bottom panel displays
the {\it Hubble Space Telescope} (HST) $R$-band image of the southern
half of this galaxy. On the southern side of the
galaxy, arms cross and
 outline an apparent optical triangle with base $12''$, height $15''$,
and apices labelled $a$ (the eastern apex), $b$ (the southern apex), 
and $c$ (the western apex).  We refer below to this
feature as the {\it triangle}. 
Sandage's peculiar arm outlines the eastern edge of
the {\it triangle}. It crosses the inner spiral arm, forming an ``X''-feature at
apex $a$; at apex $b$ it  produces another
``X''-feature. At the location marked $f$ in 
Figure\,\ref{B.Features}, it creates what looks like another ``X''-feature.
The inner spiral arm, which has a major dust lane
along its concave side, outlines the northern edge of the 
{\it triangle}, connecting apices $a$ and $c$. Another arm connects 
apices $b$ and $c$ to form the western edge of the {\it triangle}. 
At apex $b$ it meets Sandage's peculiar arm. Southward of apex $b$
the two arms diverge, and Sandage's peculiar arm extends to the west.
We call the latter the western antenna and label it $e$ in this Figure.
On the northeastern side of NGC 3145, we label as Feature $d$ the 
location where a feature that optically looks like a spiral-arm branch 
departs from the main spiral arm. We shall refer to this as the {\it branch}
even though the data presented here suggest that it is a tidal arm.

In his study of caustic waveforms in two dimensions, 
\citet{struck90} notes that the {\it triangle} 
in NGC 3145  resembles a swallowtail caustic. In a swallowtail 
caustic, five intersecting star streams create a triangular region outlined by 
arms plus two thin antennae emerging from one of the triangle's
vertices. In his study, swallowtail caustics develop within expanding 
pseudo-rings.

Gas streams do not pass through each other without colliding and shocking.
To study the above optical anomalies and look for other unusual features in the
gas, we made Karl G. Jansky Very Large Array 
 (VLA)\footnote{The National Radio Observatory is a facility of the
 National Science Foundation operated under cooperative agreement by 
 Associated Universities, Inc.}  \HI\ observations of the NGC 3145 triplet  
and \sixcm\
radio continuum observations of NGC 3145. We use the radio continuum 
image to check on whether there are strong shock fronts at the
``X''-features as would be expected if the intersecting arms are in the
same plane.

The purpose of the \HI\ data is to look for evidence of gas flows in 
extra-planar gas and gas flows perpendicular to the disk 
over large spatial scales,  e.g., are there multiple 
components in the \HI\ line profiles. 
For edge-on galaxies displacements above the disk would be obvious.
Since the inclination $i$ of NGC 3145 is $50\degr - 55\degr$, we get
only indirect information.

\begin{deluxetable*}{lccc}
\tabletypesize{\scriptsize}
%\tablenum(1)
 \tablecaption{Basic Optical and Near-Infrared Data on the 
NGC 3145 Triplet\tablenotemark{a}
\label{basic.optical}}
\tablewidth{0pt}
\tablehead{
 \colhead{Characteristic} & \colhead{NGC 3145} & \colhead{NGC 3143} 
   & \colhead{PGC 029578}
}
\startdata
Morphological type   &    SB(rs)bc     &        SB(s)b      &         Sdm\\
Right ascension (J2000)   &  $10^h 10^m 09\fs869$ &  $10^h 10^m 03\fs98$
     &  $10^h 10^m 02\fs32$\\
Declination (J2000)     & $-12\degr 26' 01\farcs6$  & $-12\degr 34' 52\farcs9$
     &   $-12\degr 38' 51\farcs6$\\
$v_{\rm sys}$(helio) (\kms)  & $3652 \pm 6$  & $3536$  &  $3586 \pm 30$ \\
Isophotal major diameter $D_{25} $  & $194'' \pm 13''$ &
       $52'' \pm 8.7''$  &  $102'' \pm 16.6''$ \\
$D_{25}$ diameter (kpc)    & 51  &  14  &  27\\
$B_T$ (RC3) (mag) \tablenotemark{b}          
    &  $12.54 \pm 0.13$ &  $14.9 \pm 0.20$  & \nodata\\
Corrected $B_T^0$   &  11.82        &  14.46    &   \nodata\\
Galactic $A_B$         &   0.234      &  0.244      &   0.25\\ 
$M_B$ (mag)        & $-21.87$   &  $-19.2$  &  \nodata\\
$B$ (Fick Obs.) (mag)\tablenotemark{b}
     &  \nodata\  & 15.2 & 16.6\\
$R$ (Fick Obs.) (mag)\tablenotemark{b}
     &  \nodata\  & 13.7 & 15.3\\
$(B - V)_T^0$          &   0.63          &    0.50      &  \nodata\\
$(K_s)_{\rm total}$  (mag)  & $8.62 \pm 0.029$  &  $11.46 \pm 0.11$
    & \nodata \\
$J_{\rm  total} $       & $9.59 \pm 0.017$ & $12.20 \pm 0.05$ & \nodata\\
$i$ from axis ratio  & $55\degr$  &  $35\degr$  & $68\degr$\\
Major Axis PA  from $K_s$      &  $205\degr$   &  $225\degr$  & \nodata\\
Projected separation from NGC 3145 (kpc) & 0  & 141 &  204\\
Distance  (Mpc)      & 54.8   & 54.8  & 54.8\\
Scale  (pc per $1''$)  & 262 & 262 & 262\\
\enddata
\tablenotetext{a} {$v_{sys}$ of  PGC 029578 from \citet{zaritsky97}
(but is not consistent with our HI value)  and
$v_{sys}$ of NGC 3143 from \citet{schweizer87}. Aside from the
Fick observatory values of $B$ and $R$ for the companions,
 the rest of the data are from \citet{deVau91} or 2MASS}.
\tablenotetext{b} {Not corrected for extinction}
\end{deluxetable*}

NGC 3145 is one of the 11 galaxies in the seminal study by 
\citet{rubin78} on extended rotation curves of high-luminosity spiral galaxies.
\citet{rubin82}  include NGC 3145 in
their study of the rotational
properties of Sb/bc galaxies from optical observations 
of the rotation curves along the major axis. 
Their data are used by 
\citet{persic91} in fitting a universal rotation curve to spiral galaxies. 
Our \HI\ observations are the first determination of the velocity
field image of NGC 3145 and the kinematic parameters derived from it. 
 It is important to know
if gas flows perpendicular to the disk have a significant effect on its rotation
curve.

From the NASA/IPAC Extragalactic Database (NED), we adopt a redshift-based 
distance for NGC 3145 of 54.8 Mpc, which is cosmology-corrected
to the 3 K microwave background reference frame with
$H_0$ = 73 kpc s$^{-1}$ Mpc$^{-1}$, $\Omega_m$ = 0.27,
and $\Omega_\lambda$ = 0.73 . Then $1''$ = 262 pc.

Table\,\ref{basic.optical} lists basic optical and near-infrared properties
of the NGC 3145 triplet from the NASA/IPAC Extragalactic Database and
includes data from the Third Reference Catalogue \citep{deVau91} and
from  the Two Micron All Sky Survey  ({\it 2MASS}) \citep{skrut06}.
In the literature, there is not very much data about 
the two companions, and the determination of their systemic velocities
has had a checkered history. Our \HI\ observations resolve the
issues about the systemic velocities of the 
companions and provide values of their \HI\ masses.
Our velocity field image of
PGC 029578 allows us to calculate its dynamical mass.

All of the velocities listed in this paper are heliocentric and use the
optical definition of the nonrelativistic Doppler shift.

 Section 2 describes our VLA observations, data reduction, and additional
images used in this paper. Section 3 discusses the broad-band optical,
\Halpha, radio continuum, and infrared properties of NGC 3145. 
Section 4 presents
the \HI\ properties of NGC 3145. Section 5 presents our \HI\
results on the two  companions. 
Section 6 summarizes and discusses our conclusions.  
A simple analytic model for the encounter is given in the Appendix.

\section{\label{observe} Observations and data reduction}

\subsection{\HI\ Observations}

We observed the NGC 3145 triplet in \HI\ at the VLA for 4.75 hr (on target)
in C configuration on 1993 July 16 and for 1 hr in D configuration on 1993
October 31. The phase calibrator was 0941-080. For the D configuration
observations 3C 286 served as the flux standard and bandpass calibrator.
For the C configuration observations 3C 48 and 3C 286 were the flux standards
and bandpass calibrators. The phase center was R.A., decl. (2000) =
10 10 06.915, -12 29 46.80.  We used on-line Hanning smoothing.
\HI\ emission is present for heliocentric velocities
3407 to 3919 \kms. 

For the correlator mode, we adopted  the same type of 4IF  (intermediate frequency) 
mode used by
\citet{elmegreen95} for \HI\ observations of NGC 2207/IC 2163. Thus we made
simultaneously (a) observations with a channel width of 21.13 \kms\ in IFC
and IFD (left circular polarization) and (b) observations with a channel width of 5.28  \kms\ in
IFA and IFB (right circular polarization).  The  line-free channels of the data taken with
21.13 \kms\ velocity resolution provide the continuum to subtract from the higher 
velocity-resolution data. The latter had no line-free channels at the low frequency end 
and only a short range of line-free channels at the high frequency end. In retrospect,
this was not the best choice of correlator mode for  the NGC 3145 observations
because of beam squint, as explained below.

The AIPS software package was used for the data reduction and analysis. The 4 IFs
plus two configurations resulted in 8 $uv$ data sets. In each of these, two channels
suffered radio frequency interference (RFI) which occurred at different frequencies
in the various data sets. To correct for this, we made a continuum image from the 
line-free channels of IFC and IFD and subtracted the clean components 
of the brighter continuum sources from every channel in each of the 8 $uv$ data sets. 
Then for the channels 
affected by RFI, we interactively clipped any signal above 1 to 2 Jy. 

For each $uv$ data set separately, we generated ``dirty'' data cubes of line plus
residual continuum emission, subtracted the residual continuum obtained from
IFC and IFD, and cleaned the cubes. Then combining the data from C and D 
configurations, we repeated the mapping and cleaning and merged the IFs to create one 
cube with 21.13 \kms\ velocity resolution and one with 5.28 \kms\ velocity
resolution. After inspection, we decided to omit the 5.28 \kms\ velocity resolution 
data taken with D configuration because beam squint, i.e., a slight difference in pointing
between the left and right polarization detectors, caused a problem for the D-configuration
data when the continuum derived from IFC plus IFD was subtracted from the 
observations made with IFA and IFB.
 Since beam squint is less of a problem for C-configuration, we kept
the C-configuration data cubes with 5.28 \kms\ velocity resolution, but to
increase the S/N we averaged the channels to 10.57 \kms\ velocity resolution.

\begin{deluxetable*}{lcccc}
\tabletypesize{\scriptsize}
%\tablenum(2)
\tablecaption{Final \HI\ Subcubes\label{HIcubes}}
\tablewidth{0pt}
\tablehead{
  \colhead{Parameter} & \colhead{Cube 1} & \colhead{Cube 2} & \colhead{Cube 3}
& \colhead{Cube 4}
 }
\startdata
 Configuration  &  C + D  &   C + D  & C & C\\
Channel width (\kms)  &  21.13  & 21.13   & 10.57  &  5.28\\
Weighting  &  Natural  & Robust 0.1 & Natural  &  Natural\\
PSF (FWHM, PA)  &  $30\farcs6 \times 18\farcs4, -31\degr$
   & $22\farcs0 \times 14\farcs7, - 28\degr$ &  $ 27\farcs3 \times 16\farcs6, - 29\degr$ 
   &  $27\farcs3 \times 16\farcs6, -29\degr$\\
Pixel Size  & $5''$  & $5''$ &  $5''$ & $5''$\\
Number of channels & 32 &  32  & 52 & 104\\
$T_b/I$ (K/\mJybeam)  & 1.10  &  1.92  &  1.37 & 1.37\\
$\sigma_{\rm rms}$ (\mJybeam)\tablenotemark{a} 
    &  0.50  &  0.58  &  0.74  &  1.04\\
\enddata
\tablenotetext{a} {rms noise per channel}
\end{deluxetable*}

To select areas of genuine \HI\ emission, we convolved the cube made with 
natural weight and 21.1 \kms\ channels 
to $60''$ resolution, clipped it at 2.5 times its rms noise, and retained regions of
emission only if the feature appeared in at least two adjacent velocity channels.
The resulting cube was applied as a blanking mask to the other cubes.
Table\,\ref{HIcubes} lists properties of our four final \HI\ data subcubes.
%used in this paper.
The one with highest sensitivity is Cube 1, made from C plus D configuration data with
natural weighting and  21.1 \kms\ velocity resolution. The one with highest 
spatial resolution is Cube 2, made from C plus D configuration data with
Robust = 0.1 in the AIPS task IMAGR and 21.1 \kms\ velocity resolution. The
cube with highest velocity-resolution (and the largest rms noise) is Cube 4 
made from C configuration data alone with natural weighting and 
5.28 \kms\ velocity resolution.  Cube 3 was made from Cube 4 by averaging
the channels to 10.57 \kms\ velocity resolution to reduce the rms noise.

For the line-of-sight \HI\ column density image displayed in the figures below,
we convolved the zeroth moment image made from Cube 2 to a 
circular $22''$ (HPBW) beam and corrected for primary-beam attentuation.
This correction factor has a mean value of  1.04 at the location of NGC 3145,
 1.07 at the location of NGC 3143, and 1.26  at the location of  PGC 029578. 

For NGC 3145, we find the integrated \HI\       
line flux $S$(HI) is 20.4 Jy \kms, which corresponds to an \HI\ mass
$M$(HI) of $1.4 \times 10^{10}$ \msun. Our value of $S$(HI) for 
NGC 3145 is 30\% greater than the \HI\ Parkes All Sky Survey value 
\citep{doyle05} of 15.4 Jy \kms.  For  NGC 3143,
 we find $S$(HI) is 0.91 Jy \kms,
and thus $M$(HI) = $6.5 \times 10^8$ \msun. As we shall see below, NGC 3143 is
somewhat deficient in \HI.  For PGC 029578, we find $S$(HI)
 is 4.7 Jy \kms, and thus $M$(HI) = $3.3 \times 10^9$ \msun.

\subsection{Radio Continuum Observations  at  \sixcm}

With the VLA, we observed NGC 3145 in the radio continuum at a central frequency of
4860.1 MHz for 1 hr (on target) in C configuration on 1994 October 21 and for 
48 min in D configuration on 1995 May 13. The observations were made with one 
IF pair at 4885.1 MHz and the other at 4835.1 MHz, each with a 50 MHz bandwidth.
The phase center was R.A., decl.(2000) =
10 10 09.995, -12 26 01.90, within $2''$ of the NGC 3145 nucleus. The phase
 calibrator was 0941-080, and the flux calibrator was 3C 286.
The smaller galaxies NGC 3143 and PGC 029578 were too far from the phase center 
to be detected.

The AIPS software was used for the data reduction.  After calibrating  the
$uv$ data 
from each VLA configuration separately and checking the separate maps, we 
combined the $uv$  data from the two configurations and ran the AIPS task IMAGR with
ROBUST = 0 to make and clean a map with synthesized beam  $7.4'' \times 5.5''$
(HPBW) and BPA = $12\degr$. A surface brightness of 1 \mJybeam\ corresponds
to $T_b$ = 1.275 K, and the rms noise $\sigma_{\rm rms}$ is 0.0224 \mJybeam, 
equivalent to $T_b$ =  0.029 K.  The mean correction for primary beam attenuation 
in NGC 3145 is a factor of 1.009.
We also convolved this image to a circular beam of $7.5''$ (HPBW). In this image,
displayed in the figures below, a surface brightness of 1 \mJybeam\ corresponds to
$T_b$ = 0.920 K and the rms noise is 0.0228 \mJybeam, equivalent to $T_b$ = 
 0.021 K.
For NGC 3145 we find a total flux density $S_\nu$(4.86 GHz) = $9.1 \pm 0.3$  mJy.

\subsection{Additional Data}

We use \Halpha, broad-band optical, and infrared images that other observers
and facilities have made available on-line. Table\,\ref{PSF} lists the FWHM 
of the point-spread functions  (PSFs) of  these images.
The continuum-subtracted \Halpha\ image of NGC 3145 is from 
\citet{banfi93} and is not flux-calibrated. The $B$ image in 
 Figure\,\ref{B.Features} is from \citet{sandage94}.
Whenever we refer to a $B$ image of NGC 3145 without further specification, 
we  mean this image.  We use the other $B$ images described below
when we need a larger field.  We obtained a WFPC2, F606W ($R$-band) 
Hubble Space Telescope ({\it HST}) image from 
the {\it Hubble Legacy Archive}\footnote{Based on observations made 
with the NASA/ESA Hubble Space Telescope, and obtained from the 
Hubble Legacy Archive, which is a collaboration between the Space
Telescope Science Institute (STScI/NASA), the Space Telescope 
European Coordinating Facility (ST-ECF/ESA) and the Canadian 
Astronomy Data Centre (CADC/NRC/CSA)}.
 \citet{martini03} made these {\it HST} observations to study 
circumnuclear dust, but in addition to the central regions, 
the field covers the southern half of NGC 3145.

\begin{deluxetable}{lc}
\tabletypesize{\scriptsize}
%\tablenum(3)
\tablecaption{Resolution of Images\tablenotemark{a}
\label{PSF}}
\tablewidth{0pt}
\tablehead{
  \colhead{Image}  &  \colhead{FWHM of PSF}
}
\startdata
Carnegie Atlas of Galaxies $B$       &  $1.3'' \times 1.2''$\\
{\it MDM}  $B$                                &  $2.3'' \times 1.9''$\\
{\it MDM} $V$                                 &  $1.9'' \times 1.6''$\\
{\it MDM}  $R$                                &  $1.7'' \times 1.5''$\\
Burrell-Schmidt $B$                       &  $4.1''$\\
{\it HST} WFPC2 $R$                       &  $0.20'' \times 0.18''$\\
DSS IIIaJ                                          &   $5.4'' \times 4.4''$\\
\Halpha\                                        &  $\sim 1''$\\
\sixcm\ radio continuum               &  $7.4'' \times 5.5''$\\
\sixcm\ radio continuum               &  $7.5''$\\
$N$(HI)                                        &   $22''$\\
{\it WISE} 12 \micron\                    &  $9.2'' \times 8.4''$\\
{\it WISE} 22 \micron\                   &  $18'' \times 17''$\\
\enddata

\tablenotetext{a} {Except in the case of the radio images and
the \Halpha\ image,
the FWHM values are from fitting  two-dimensional 
Gaussians to stellar images in the optical or WISE 3.4 \micron\
image. For the \Halpha\ image, we list the value of the estimated
seeing from \citet{banfi93}. 
}
\end{deluxetable}

We obtained  from the NASA/IPAC Infrared Science Archive
 infrared images of NGC 3145 
taken by the Wide-field Infrared
Survey Explorer ({\it WISE}). We also use data from 
{\it 2MASS}.

 Paul Eskridge took $B$, $V$, and $R$ images of NGC 3145
for us at the 1.3 m  telescope of the Michigan-Dartmouth-M.I.T. Observatory
(MDM)  on 1999 March 25.  These images, taken under non-photometric 
conditions, are a little trailed and a little underexposed.
We took a $B$ image with the Burrell-Schmidt telescope at
Kitt Peak on 1993 January 29 and use this for NGC 3143.
 These images were bias subtracted, 
flat-fielded, combined, and sky-subtracted with standard IRAF procedures.

For PGC 029578 we use a IIIaJ (4680 \AA) Digitized Sky Survey (DSS) 
image from Space Telescope Science Institute.

Philip Appleton took $B$ and $R$ images for us of NGC 3143 and
PGC 020578 at the 0.6 m telescope of  the Fick Observatory of
Iowa State University, and he measured the (Johnson) $B$ and $R$ 
magnitudes  listed in Table\,\ref{basic.optical} for these two galaxies.

\section{Broad-band Optical, \Halpha, Radio Continuum, and Infrared 
Properties of NGC 3145}

Throughout this section we use for deprojection of  NGC 3145
the values of the position angle (PA) of the major axis and the inclination $i$ of 
the disk from Table\,\ref{basic.optical}. Our \HI\ kinematic  data in Section 4.2 find a
value for the major axis PA consistent with the isophotal value in this table and  a
somewhat smaller value of  $i$. 

\subsection{Broad-band Optical}

\begin{figure*}
\epsscale{0.9}
\plotone{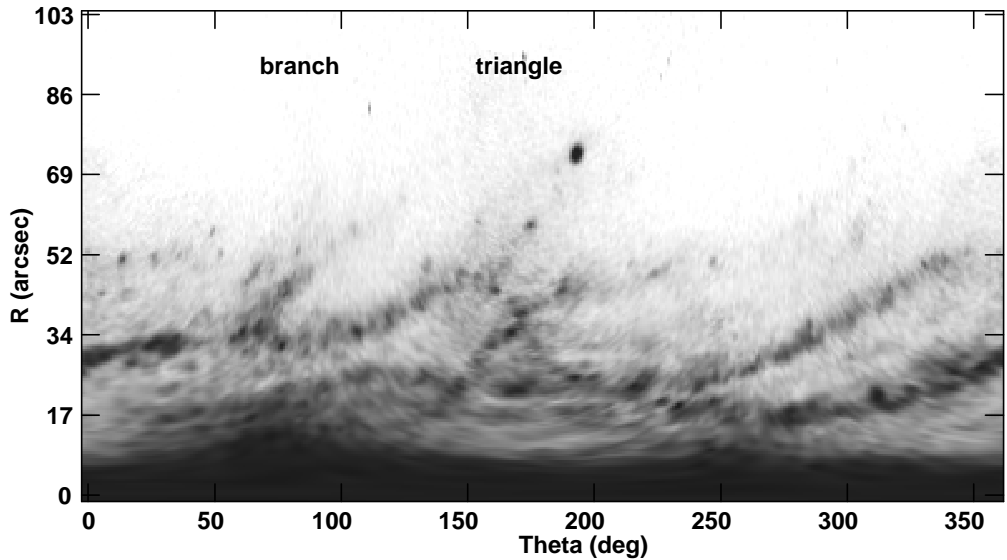}
\caption{Greyscale polar coordinate display of the $B$ image of
Figure\,\ref{B.Features} after deprojection into the plane of 
NGC 3145. The azimuth $\theta$ is measured ccw from the NNE
major axis.
Some notable features are the {\it triangle} and
the apparent {\it branch} of the main spiral arm
starting at Feature $d$, as marked in Figure\,\ref{B.Features}.
The labels ``branch" and ``triangle"  indicate the 
approximate values of $\theta$ for
these features. Apex $a$ of the {\it triangle} has $\theta \simeq 146\degr$.
The arm connecting apices $b$ and $c$ of the {\it triangle} winds 
in opposite sense to all the other arms.
\label{az}}
\end{figure*}

The {\it HST} $R$-band image of the southern half of NGC 3145, displayed
in the bottom panel of Figure\,\ref{B.Features}, 
 affords a detailed view of some of the optical oddities  marked in the
top panel of that figure.
The larger box encloses the {\it  triangle}, and the smaller box is on
the western antenna. Sets of \HI\ line-profiles from these two boxes will
be displayed and discussed in Section 4.2.2. Heading north from
apex $c$ of the {\it triangle} is a region of complex dust
loops and feathers. 
{\it HST} images of Sb and Sc galaxies often reveal dust feathers crossing
 spiral arms \citep{la vigne06}. Numerical
 models of spiral galaxies by \citet{kim02}, \citet{kim06}, and \citet{shetty06}
reproduce feathers and spurs as sheared structures resulting from magneto-Jeans
instabilities as gas flows through a spiral shock, and \citet{lee12} and \citet{lee14}
present an analytic discussion of the feathering instability of spiral arms
for a thin disk with magnetic fields and self-gravitating gas.
The dust feathers in this region
of NGC 3145 are spaced 0.3 to 0.7 kpc apart and are probably produced by
this mechanism. It is not clear whether this mechanism accounts for  
the dust loops in this part of NGC 3145. We shall refer to this region 
as the region of complex dust loops. 

Sandage's peculiar arm, which outlines the eastern edge of the {\it triangle},
is prominent in the HST image.  Crossed by a number of dust feathers, it
has a dust lane along part of its outer edge and another along
part of its inner edge. 
This arm looks to be in front of the inner spiral arm, which it crosses at apex $a$,
and thus in front of the disk. Figure\,\ref{az} is a display of the
$B$ image in polar coordinates after deprojection into the 
plane of NGC 3145. In this image,
the {\it triangle} is a prominent feature,  and Sandage's peculiar arm has 
quite a different slope than the spiral arms, consistent with it being a
material arm in front of the disk. The {\it branch} starting 
at Feature $d$ is also visible in this image. Its slope in this 
polar-coordinate diagram is about the same as that
of Sandage's peculiar arm between apices $a$ and $b$.

The arm which
outlines the western edge of the {\it triangle}  and connects
apices $c$ and $b$ winds in opposite sense to every other 
arm in NGC 3145 (see Figures\,\ref{B.Features} and 
\,\ref{az}); it increases in radius cw whereas all the other arms
increase in radius ccw. It is  less
prominent in the $R$-band {\it HST} image
than in the $B$ image of  Figures\,\ref{B.Features} and \,\ref{az}. 
In our MDM images of NGC 3145,
this  arm decreases in prominence from $B$ to $V$ to $R$.
This suggests it is dominated in $B$ by somewhat younger stars.

In Figure\,\ref{B.Features} there is also a string of bluish clumps crossing
Sandage's peculiar arm at Feature $f$ ($6''$ north of apex $a$) and
continuing,  quasi-parallel to the northern edge of the {\it triangle},
to the southern part of the region  of compex dust loops. 
 This string of clumps is not evident in the {\it HST} image. 

\begin{figure*}
\epsscale{1}
\plottwo{fig5.eps}{fig6.eps}
\caption{Left: Greyscale plus contour display of the \sixcm\ radio 
continuum image of NGC 3145 with $7.5''$ resolution. Contour
levels are at (3, 5, 6, 7, 9) $\times$ the rms noise. The rms noise
is 0.0228 \mJybeam, equivalent to $T_b$ = 0.021 K. See text about
Feature $f$.
Right: Greyscale display of \Halpha\ image overlaid with contours
of \sixcm\ radio continuum emission. Several of the \sixcm\ radio knots
coincide with \Halpha\ clumps. The cross
marks the location of the nucleus.
\label{6cm}}
\end{figure*}

\begin{figure*}
\epsscale {0.7}
\plotone{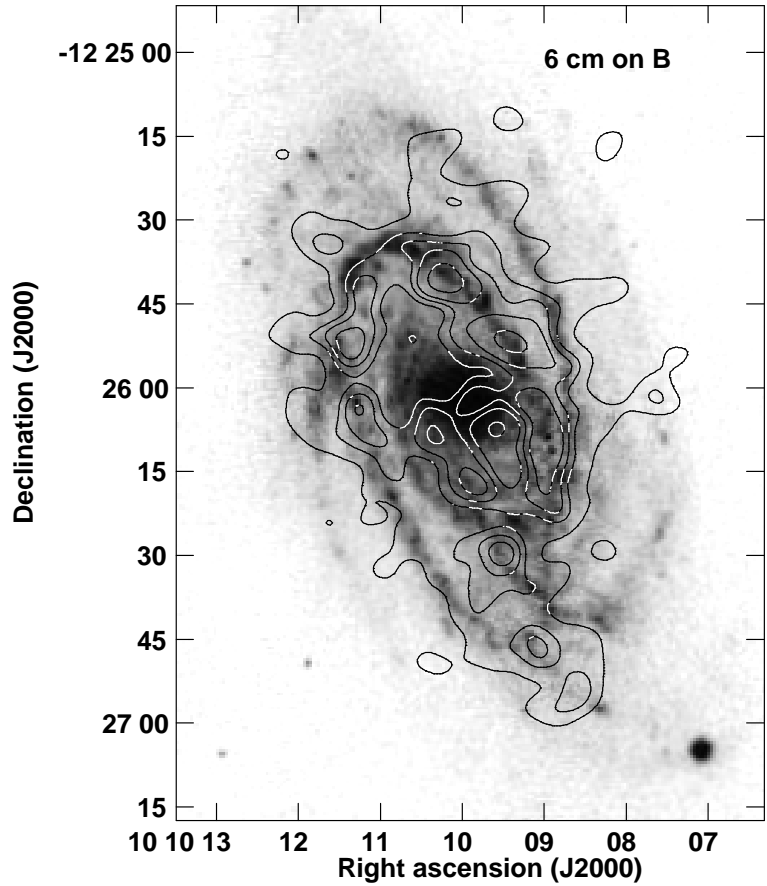}
\caption{Greyscale display of $B$ image of NGC 3145
overlaid with contours of \sixcm\ radio emission from
Figure\,\ref{6cm}. {\it Apices $a$ and $b$ of the triangle
are not prominent in the \sixcm\ radio.}  North of Feature $f$
(marked in Figure\,\ref{6cm}) there
is extended \sixcm\ radio emission along Sandage's peculiar 
arm.
\label{B.6cm}}
\end{figure*}

From {\it 2MASS} $K_s$ isophotes, the position angle (PA) of the major
axis of the bar is 60\degr. The region of complex dust loops and the location
where the {\it branch} (feature $d$ in Figure\,\ref{B.Features}) departs
 from the spiral arm are on opposite sides of the galaxy  at somewhat
different distances from the nucleus and
roughly along the same PA as the major axis of the bar.

\subsection{Radio Continuum, \Halpha, and mid-IR}

For NGC 3145, the NRAO/VLA Sky Survey (NVSS) \citep{condon98} lists 
a total flux density $S_\nu$(1.4 GHz) of $21.5 \pm 2.3$ mJy.
Since we find $S_\nu$(4.86 GHz) = $9.1 \pm 0.3$ mJy,
the global spectral index $\alpha$ of NGC 3145 =
$-0.7 \pm 0.1$ , which is typical of normal spirals.  For
normal galaxies with a spectal index in this range, the expression in
\citet{condon92} for the ratio of free-free to total flux density gives
a thermal fraction at \sixcm\ for a galaxy as a whole
of 20\% to 25\%.

Usually in spiral galaxies, the extended emission along the spiral arms 
is predominantly nonthermal, and only where knots occur is the
emission mainly free-free.
Young star-forming complexes appear as \Halpha\ sources and knots
 of  \sixcm\ radio continuum emission in NGC 3145.
The \sixcm\ radio continuum is also used to search for possible shock fronts.
 The left panel in Figure\,\ref{6cm} displays in greyscale and 
contours the \sixcm\ radio 
continuum image with $7.5''$ (= 2.0 kpc) resolution. In the right panel,
these contours
are overlaid on a greyscale display of the \Halpha\ image 
(which has a resolution of $\sim 1''$ = 260 pc).
 In Figure\,\ref{B.6cm}, the \sixcm\ contours are overlaid
 on the $B$ image in greyscale to show where the radio continuum emission is 
located relative to the features marked in Figure\,\ref{B.Features}.

The \sixcm\ knots tend to lie along the brighter parts of the $B$-band
spiral arms, and several coincide with \Halpha\ clumps, 
but not all of the brighter \Halpha\ clumps are seen as \sixcm\ 
emission knots.

The two brightest \sixcm\ clumps lie on the inner
and brighter of the two $B$-band spiral arms in the north 
and have flux densities $S_\nu$(6 cm) = $0.44 \pm 0.04$ mJy
for the northwestern clump and $0.41 \pm 0.04$ mJy for the 
northern clump. Neither of these is at the ``X''-features 
marked in Figure\,\ref{B.Features}.

\begin{figure}
\epsscale{0.70}
\plotone{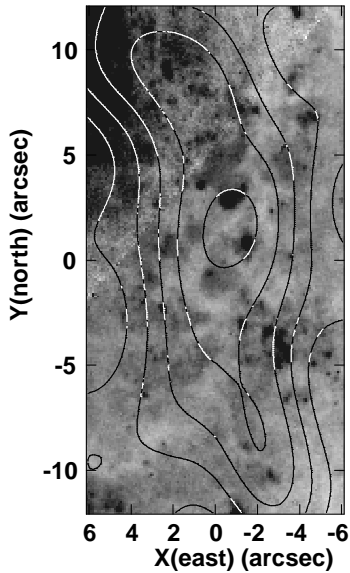}
\caption{In greyscale the portion of the {\it HST}
image containing the region of complex dust loops
and centered on R.A., decl.(2000) = 10 10 09.006, 
$-12$ 26 11.80.
This is overlaid with contours at 
(3, 4, 5, 6, 7) $\times$ the rms noise
from the \sixcm\ radio continuum image
with $7.4'' \times 5.5''$ (HPBW) resolution. The rms noise
is 0.0224 \mJybeam, equivalent to $T_b$ = 0.029 K.
Some of the extended \sixcm\ emission here may be from
shocks associated with the dust loops and feathers.
\label{HST.6cm}}
\end{figure}

In Figure\,\ref{HST.6cm}, contours from our \sixcm\ radio
continuum image with higher resolution ($7.4'' \times 5.5''$ HPBW) 
are overlaid on the portion of the $HST$ image containing the 
region of complex dust loops.
 This region has \sixcm\ flux density $S_\nu$(6 cm) =
$0.73 \pm 0.06$ mJy. Although there is \Halpha\
emission here, some of the extended \sixcm\ 
emission may be from shocks associated with the 
dust loops and feathers.
The \sixcm\ clump centered where a large dust loop or shell seems 
to cross another feature here
has $S_\nu$(6 cm) = $0.21 \pm 0.03$ mJy.

\begin{figure*}
\epsscale{0.9}
\plottwo{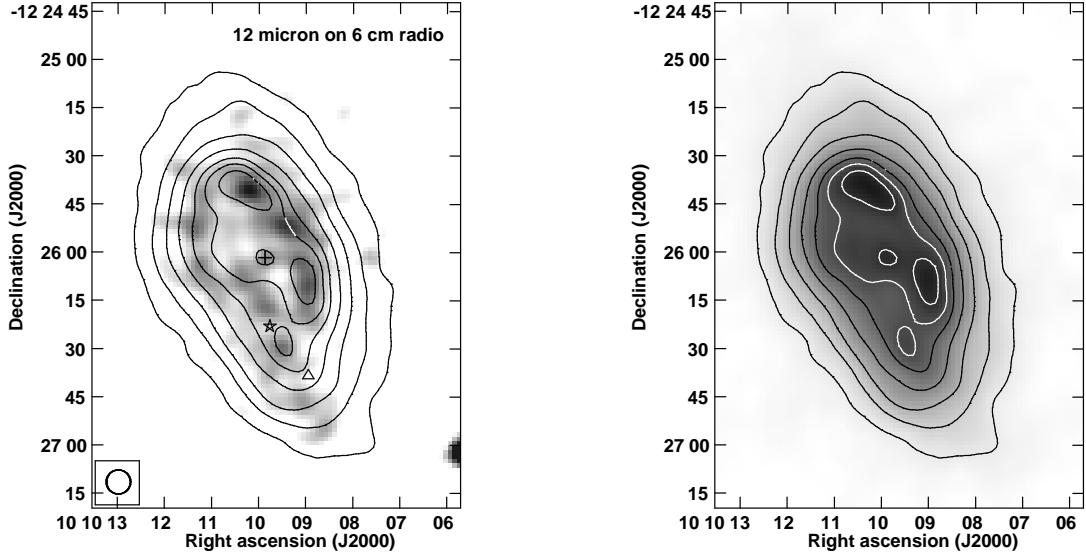}{fig9b.eps}
\caption{Left panel: Contours of {\it WISE} 12 \micron\ emission 
overlaid on \sixcm\ radio continuum in greyscale. The 
locations of
the nucleus, apex $a$, and apex $b$ of the {\it triangle}
are denoted by a plus sign, a five-pointed star, and a small triangle 
symbol, resp. Right panel: Greyscale plus contour display
of the {\it WISE } 12 \micron\ image.
Unlike the \sixcm\ radio emission,
the distribution of  12 \micron\ emission 
 is not somewhat ring-like.
\label{12micron}}
\end{figure*}

The following properties of Sandage's peculiar arm, the
{\it triangle}, and the ``X''-features are apparent from a
comparison of 
Figures\,\ref{6cm},\, \ref{B.6cm}, and \,\ref{B.Features}. There is a
radio continuum knot at Feature $f$ (labelled in
Figure\,\ref{6cm}), which is where the string of blue clumps
 intersects Sandage's peculiar arm. 
North of Feature $f$ there is extended \sixcm\ radio continuum
emission  along Sandage's peculiar arm.  This 
is probably mainly synchrotron emission from shocked gas, since
extended radio emission from arms in galaxies
is usually nonthermal.  We interpret this \sixcm\ emission feature
to mean that the portion of Sandage's arm from Feature $f$
northwards is in the disk. The \Halpha\ emission,
including the Feature $f$ knot, is faint, perhaps due
to extinction, but the  12 \micron\ emission 
(see Figure\,\ref{12micron}) from this general area is 
relatively bright.  The \sixcm\ radio emission here contrasts
with that from the {\it triangle}. Aside from an
 \Halpha\ and  \sixcm\ radio continuum star-forming clump
between apices $a$ and $b$ on Sandage's peculiar arm,
the {\it triangle}, including apices
$a$ and $b$, is not prominent in the \sixcm\ radio 
continuum.  Emission at \sixcm\ from the region of complex
dust loops partly overlaps apex $c$, but the latter does not
appear as a distinct clump at \sixcm.  We conclude that  
{\it there are no shocks at the arm-crossing ``X'''s of the
 triangle, and thus the arms that appear to cross 
at the apices of the triangle must be 
in different planes.}

Thus, whereas the part of Sandage's peculiar arm from
Feature $f$ northwards appears to be in the disk, the
part of this arm that forms the eastern side of the {\it triangle} 
is not in the disk.

 The distribution of the brighter \sixcm\ emission in NGC 3145
has a somewhat ring-like appearance, composed of the 
brighter portions of the $B$-band spiral in the northern half 
of the galaxy, the section of Sandage's peculiar arm north of
Feature $f$, and  connections between these 
(see Figures\,\ref{6cm} and \,\ref{B.6cm}). 
This contrasts with the distribution of 12 \micron\
emission.
In Figure\,\ref{12micron}, contours of  the {\it WISE} 
12 \micron\ emission are overlaid on the \sixcm\ radio image 
in greyscale.
The locations of the nucleus, apex $a$, and apex $b$ are
denoted by a plus sign, a five-pointed star, and a small triangle 
symbol, resp. The brightest 12 \micron\ sources are at the locations
of a) the northern \sixcm\ clump on the spiral arm north of
the nucleus, b)  the region of complex dust 
loops, and c) the nucleus. Aside from these three sources, the
brightest 12 \micron\ emission is from the central hole in the
distribution of \sixcm\ emission.  
Similarly, with the {\it WISE} 22 \micron\ image, whose resolution
is a factor of 2 worse than the 12 \micron\ image, 
the brightest emission is from the central
hole in the \sixcm\ distribution and from the 
 northern \sixcm\ clump on the spiral arm. Thus there is
a lot of warm dust in the inner disk/bar/bulge of NGC 3145
but not a similar concentration of cosmic-ray electrons.

  In Section 3.1 we noted that the arm outlining the western edge
of the {\it triangle}  decreases in prominence
from $B$ to $V$ to $R$.  Since this arm is not prominent in
\Halpha\ either, it appears that optical
emission from this arm may be dominated by $A$ and/or  $B $ stars.
This may help constrain the age of this feature.

Feature $d$ in Figure\,\ref{B.Features} marks where the 
{\it branch} departs from the main spiral arm on the 
northeastern side of NGC 3145. Since there is no string 
of \Halpha\ knots along the {\it branch}, it does not
meet the definition of  a spur as given by  \citet{la vigne06}.
A clump prominent in \Halpha\ and \sixcm\  emission lies on  
the main spiral arm near this point.  In Section 4.2.2, we shall see
that there is a large difference in velocity between the spiral arm and
the {\it branch} at feature $d$.

\section{\HI\ Properties of NGC 3145}

\subsection{\HI\ Images}

\begin{figure*}
\epsscale{0.747}
\plotone{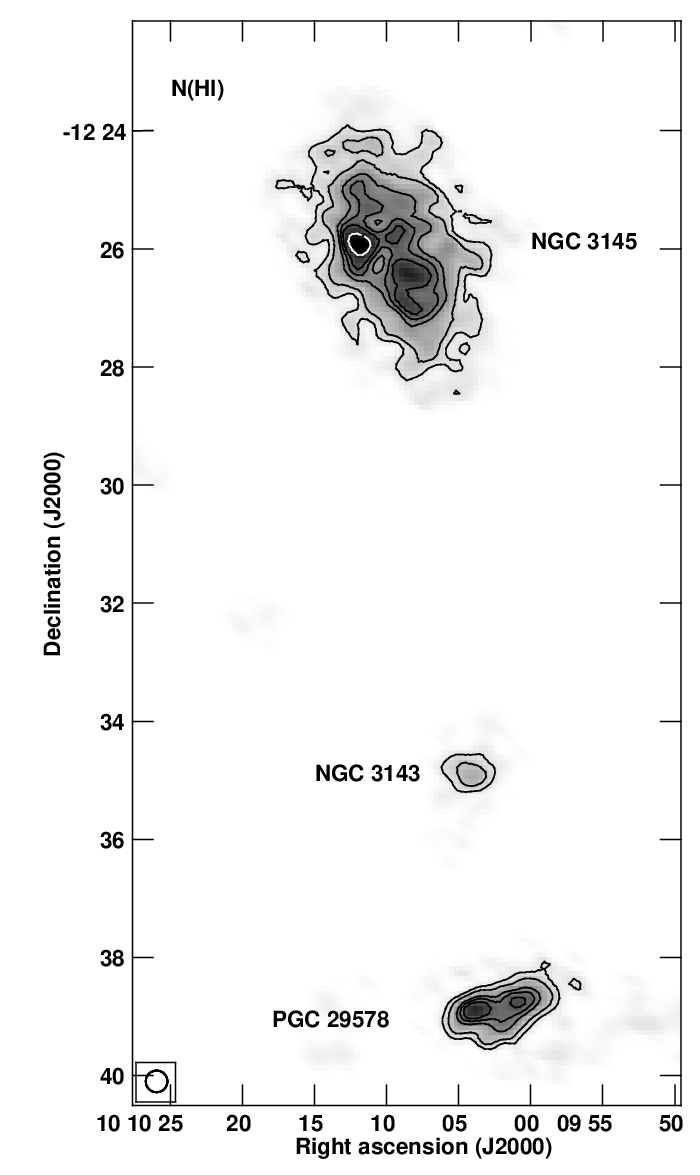}
\caption{Greyscale plus contour display of the $N$(HI)
image of  the NGC 3145 triplet with resolution = $22''$.
Contour levels are at 100, 200, 400, 500, 600, and 800 
\Jybeamms, where 100 \Jybeamms\ corresponds to a 
line-of-sight column density $N$(HI) =
 $2.3 \times 10^{20}$ \atc. The \HI\ 
distribution in NGC 3145 is not axisymmetric; there is a 
trough at $6''$ to $26''$ southeast of the nucleus and
three major \HI\ concentrations (NE, SW, and NW of the 
nucleus.) 
\label{HI.triplet}}
\end{figure*}

Figure\,\ref{HI.triplet} displays the line-of-sight column density $N$(HI) 
image of the three galaxies in the system in 
greyscale and contours 
with $22''$ resolution. Table\,\ref{HIproperties} lists the basic \HI\ 
properties of the NGC 3145 triplet from our observations.

\begin{figure*}
\epsscale{0.68}
\plotone{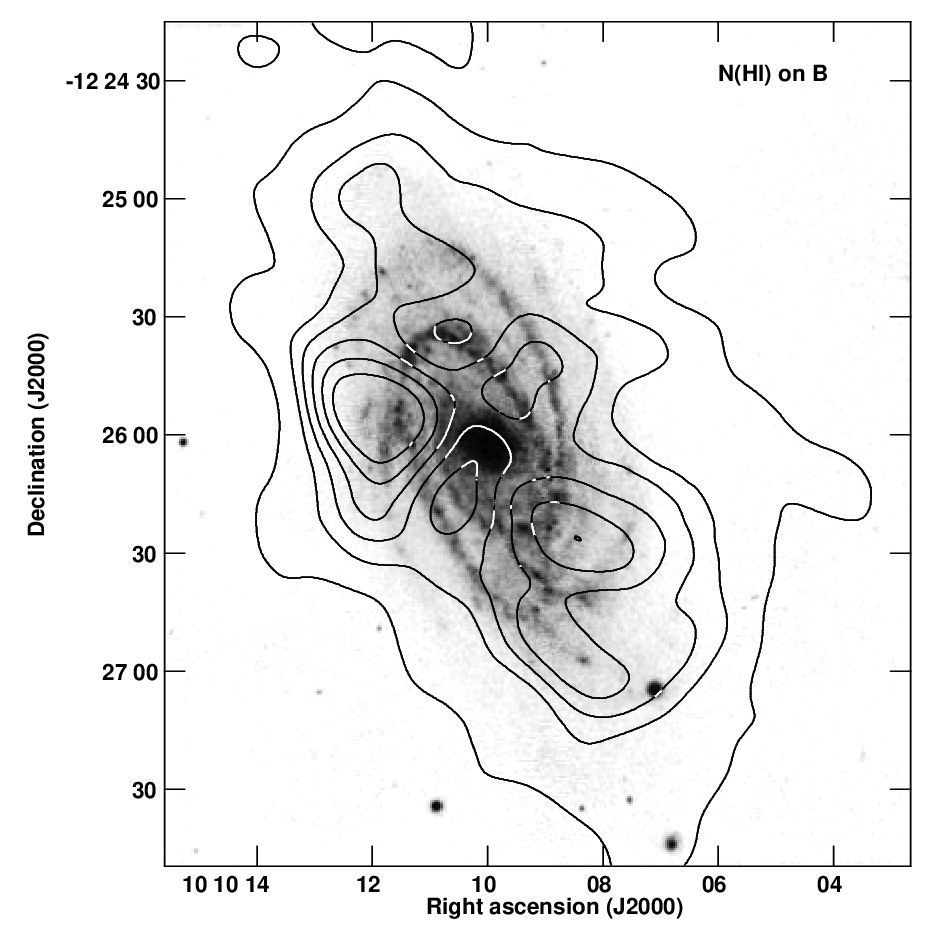}
\plotone{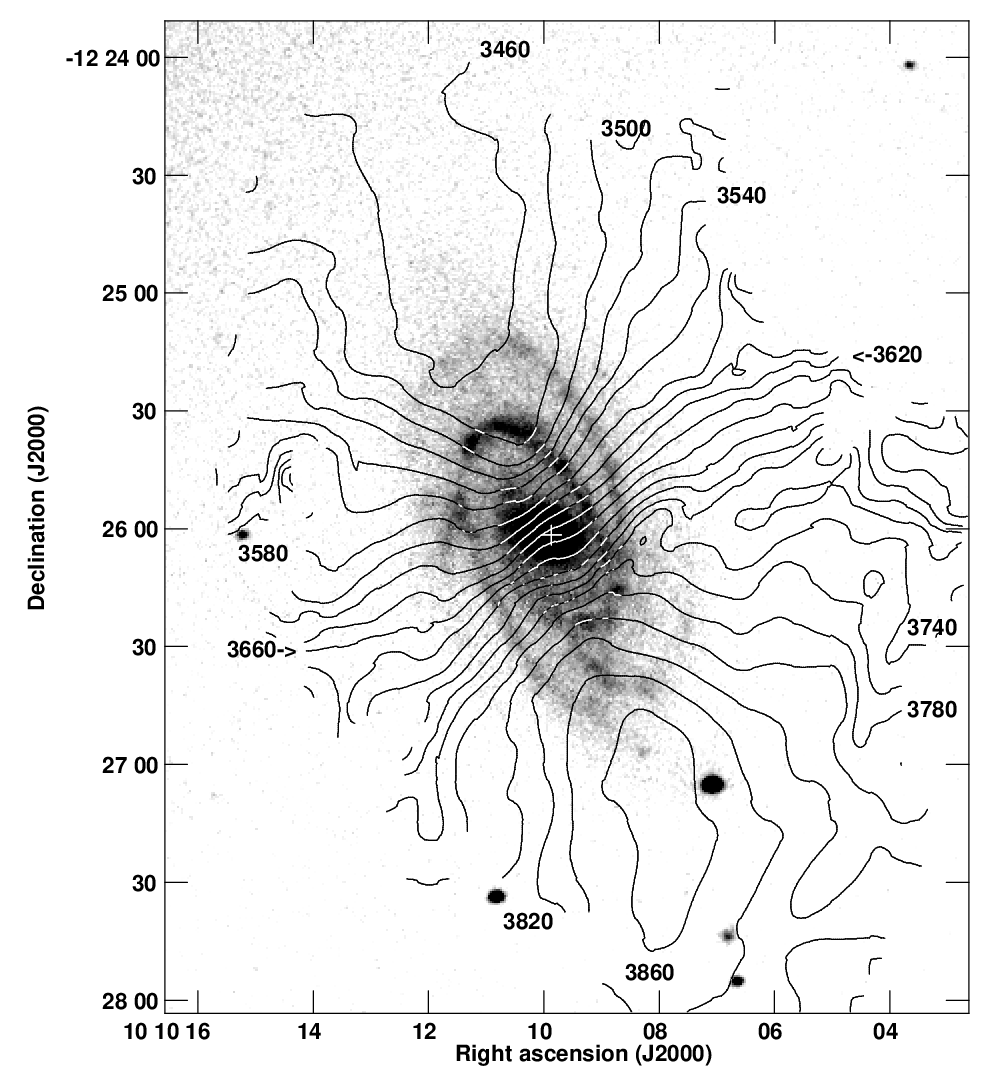}
\caption{Top: $B$ image of NGC 3145 in greyscale overlaid with
$N$(HI) contours. Contour levels are at 200, 400, 500, 
600, 700, 800 \Jybeamms, where 100 \Jybeamms\ 
corresponds to $N$(HI) = $2.3 \times 10^{20}$ \atc. 
The major \HI\ concentration northeast of the nucleus is elongated 
to the south, not along the main spiral arm but along the
{\it branch} or a bit to the east of it.
Bottom: Contours of the first moment image of NGC 3145 
overlaid on the {\it MDM} $B$ image in greyscale,
 with a plus symbol at
the nucleus.  The contour interval is 20 \kms.  Along the {\it branch}
there are kinks in the isovelocity contours.
\label{N(HI).N3145}}
\end{figure*}

\begin{deluxetable*}{lccc}
\tabletypesize{\scriptsize}
\tablecaption{Basic \HI\ Properties of the NGC 3145 Triplet
    \label{HIproperties}}
\tablewidth{0pt}
\tablehead{
   \colhead{Characteristic} & \colhead{NGC 3145} & \colhead{NGC 3143}
     &\colhead{PGC 029578}
}
\startdata
Velocity Range of HI emission  (\kms)  &  3407 to 3919  &  3491 to 3581
    &  3407 to 3623\\
Heliocentric $v_{\rm sys}$ (\kms)  & $3655.9 \pm 0.2$ &  $3530 \pm 5$
    & $3512 \pm 5$\\
PA of receding major axis:\\
    Kinematic     &   $205.44\degr \pm 0.06\degr$  &  \nodata  
         & $270\degr \pm 3\degr$\\
    Isophotal     &  $204\degr \pm 2\degr$  & $235\degr \pm 5\degr$
        &  $283\degr \pm 5\degr$\\
PA of kinematic minor axis  &  $-60\degr \pm 2\degr$  &   $\sim 0\degr$
        &  $173\degr \pm 3\degr$\\
Kinematic $i$   &  $50.3\degr \pm 0.3\degr$ & \nodata & \nodata\\
Isophotal  $i$   &   $55\degr \pm 1\degr$  & $\leq 35\degr$
         & $61\degr \pm 2\degr$\\
HI diameter  & $262'' \pm 2''$ & $52'' \pm 2''$ & $122'' \pm 2''$\\
HI diameter/$D_{25}$  & $1.35 \pm 0.09$ & $1.00 \pm 0.17$ 
          & $1.20 \pm 0.20$\\
Integrated HI flux $S$(HI)  (Jy \kms)  &  20.4  &  0.91  &  4.72 \\
$M$(HI)   (\msun)  &  $1.44 \times 10^{10}$  &  $6.5 \times 10^8$  
        &   $3.33 \times 10^9$\\
Dynamical mass $M_{\rm dyn}$ (\msun)\tablenotemark{a}
      & $5.0 \times 10^{11}$ &  \nodata  & $2.2 \times 10^{10}$ \\
$M$(HI)/$M_{\rm dyn}$\tablenotemark{a}
      &   0.029 & \nodata &  0.13  \\
Maximum HI column density  (\atc) & $2.3 \times 10^{21}$ &
     $7.0 \times 10^{20}$ & $1.8 \times10^{21}$\\
\enddata

\tablenotetext{a} {out to $R$ = $127''$ = 33 kpc for NGC 3145, and 
    $R$ = $50''$ =13 kpc for  PGC 029578}
\end{deluxetable*}

\begin{figure*}
\epsscale{0.8}
\plottwo{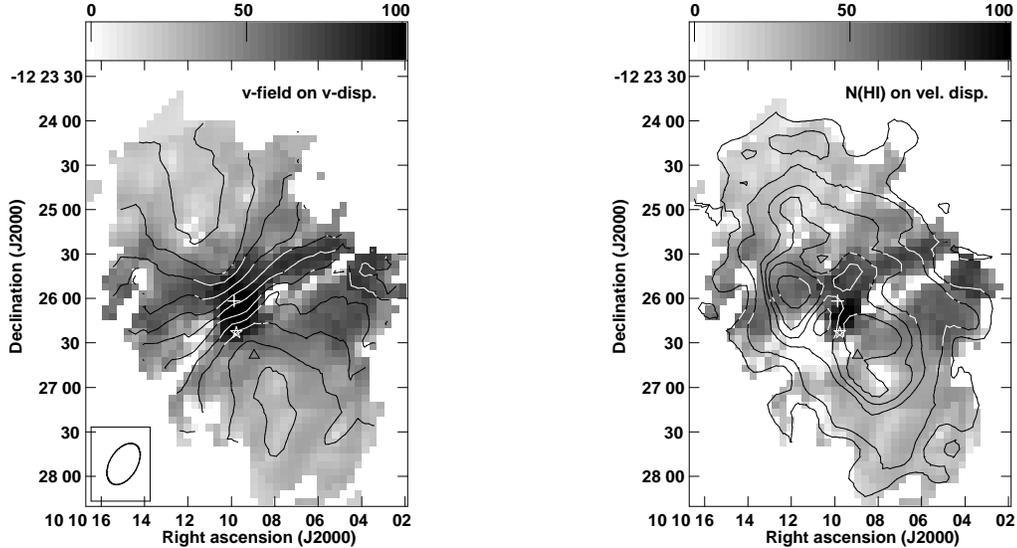}{fig13b.eps}
\caption{Left: Velocity field contours overlaid on a greyscale 
display of the velocity dispersion image before correction 
for the velocity gradient across the beam.
Right: $N$(HI) contours of NGC 3145
from Figure\,\ref{HI.triplet}
overlaid  on greyscale display of the velocity dispersion image
corrected for the velocity gradient. In both panels the units
on the greyscale wedge are \kms, and the locations of the
nucleus, apex $a$ and apex $b$ of the {\it triangle} are
marked by the plus symbol, the five-pointed star, and the 
small triangle symbol, resp. Blanked pixels are white.
\label{vel.disp}}
\end{figure*}

In the top panel of Figure\,\ref{N(HI).N3145} 
$N$(HI) contours are overlaid on the $B$ image
of NGC 3145 in greyscale.  The distribution of \HI\ emission from NGC 3145
is not axisymmetric; it consists of  three major concentrations plus fainter extended 
emission and  a trough $6''$ to $26''$ southeast of the nucleus.
  The brightest \HI\ concentration is northeast of the nucleus and
 includes Feature $d$. This region has an \HI\ mass $M$(HI) =
$1.8 \times 10^9$ \msun. It is  elongated to the south, not
along the main spiral arm but along the {\it branch} or a bit to the
east of it. At $20''$ south of  Feature $d$, the \HI\ emission from
the {\it branch} is still bright, whereas the \HI\ emission from the
spiral arm is near  the trough.
We conclude in Section 4.2.2  that this major \HI\ 
concentration  cannot be a single entity. 
The second brightest \HI\ concentration is southwest of
the nucleus
%, includes most of  the {\it triangle} and the western antenna,
 and has $M$(HI) = 
$1.4 \times 10^9$ \msun\ if we omit its southern tail along the
spiral arm.
The third brightest \HI\ concentration lies on the northwestern part of 
the two northern spiral arms,  and its brightest part
coincides with the most luminous \sixcm\ clump in the galaxy.
Its $M$(HI) = $6 \times 10^8$ \msun.  These \HI\ masses are comparable
to those of massive \HI\ clouds in a number of interacting galaxy pairs -
see \citet{kaufman99} and references therein. This suggests that the massive
\HI\ concentrations in NGC 3145 could have resulted from an encounter.

The total $M$(HI) of NGC 3145 is $1.4 \times 10^{10}$ \msun.
Its ratio of $M$(HI)/$L_{\rm B}$ = 0.17 \msun $L_{\sun}^{-1}$, which is
close to the median value for galaxies of the same Hubble type  (Sbc) in
\citet{roberts94}.

The NGC 3145 first moment image (often called the velocity field) is
displayed as contours overlaid on the {\it MDM} $B$ image in greyscale 
in the bottom panel of Figure\,\ref{N(HI).N3145}. 
The velocity dispersion image, uncorrected
for the velocity gradient across the synthesized beam, is displayed in
greyscale with velocity-field contours overlaid in the left panel
of Figure\,\ref{vel.disp}, and the velocity dispersion image after
correction for the velocity gradient is displayed in greyscale with
$N$(HI) contours overlaid in the right panel of Figure\,\ref{vel.disp}. The
first moment and velocity dispersion images  are from the cube with
lowest rms noise (Cube 1) and are blanked where
$N$(HI) $\leq 1.9 \times 10^{20}$ \atc.
A plus symbol marks the location of the nucleus. In the velocity-dispersion
figures (and in other figures below), the locations of apex $a$ and
 apex $b$ of the {\it triangle} are denoted by the
five-pointed star and the small triangle symbol, resp. 

The  velocity field exhibits some mild irregularities but not the
highly distorted velocity fields found in some 
interacting galaxies, e.g. NGC 2207 \citep{elmegreen95} or
NGC 2535 \citep{kaufman97}. Along the {\it branch}
there are kinks (or wiggles) in the velocity contours. 
At the arms where the \sixcm\ radio emission is
prominent in Figure\,\ref{6cm}, there is no evidence from 
Figure\,\ref{N(HI).N3145}, or from one
made from the C-array data with 10 \kms\ velocity resolution, of 
kinks (or wiggles) in the isovelocity contours due to streaming motions.
  If the radial component of
streaming motions dominates, such kinks should be
{\bf u}-shaped on the western side and {\bf n}-shaped on
the eastern side of the galaxy.
Either the somewhat ring-like distribution of \sixcm\ emission 
is not expanding or our \HI\ 
observations have too low a spatial resolution to reveal an expansion. 
A resonance ring produced by a bar
would not be expanding.

The velocity dispersion image that is corrected for the velocity gradient across
the synthesized beam has particularly high values for the velocity dispersion 
\sigmav\ at apex $a$ of the {\it triangle} and in the region of
the complex dust loops. For the oddities labelled in Figure\,\ref{B.Features},
 Section 4.2.2 compares numerical values of  \sigmav\ before and after 
this correction. High values for the corrected \sigmav\ may result, in part,
because the line-of-sight intercepts gas at various altitudes above the disk
and at various radial distances, in addition to streaming motions on small
spatial scales and turbulence within the disk.

\subsection{Analysis of \HI\ Motions}

\subsubsection{Large Scale Anomalies}

We use the AIPS program GAL to fit a model rotation curve 
to the first moment image of NGC 3145
and determine its kinematic parameters.
The routine assumes a flat disk in rotation. The residuals from such a 
fit may reveal whether significant  parts
of the the galaxy are not in a flat disk or have motions other than rotation.
We first used GAL to fit a Brandt model \citep{brandt60} rotation curve 
to our first moment image
 of the disk as a whole. It finds a kinematic center within $1''$ of the
optical nucleus, a systemic velocity $v_{\rm sys}$ = $3655.9 \pm 0.2$ \kms, 
close to
the optical value of $3652 \pm 6$ \kms, and a position angle of the
receding major axis  PA = $205.4\degr \pm 0.06\degr$, consistent with the
isophotal value  of 205\degr\ from  {\it 2MASS} $K_{\rm s}$
and from the $N$(HI) outer isophotes. 

\begin{figure*}
\epsscale{0.68}
\plotone{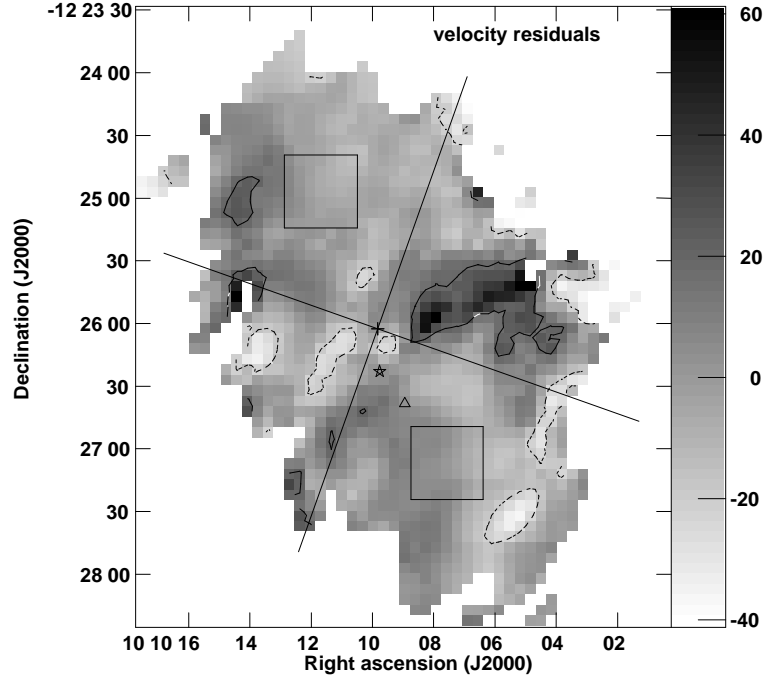}
\caption{Greyscale and contour display of velocity residuals from
fitting a Brandt model rotation curve to the first moment image
of NGC 3145 as a whole. The contours are at $20$ \kms\ and
$-20$ \kms, and the units on the greyscale wedge are
\kms. The locations of the nucleus, apex $a$ and apex $b$ 
are marked as in Figure\,\ref{vel.disp}.
 Lines are drawn to divide the galaxy
into four quadrants: receding major axis (SSW), near-side
minor axis (WNW), approaching major axis (NNE), and 
far-side minor axis (ESE).   The boxes on the major axis are 
the $R_{\rm max}$ boxes where we fit the line-profile at
each pixel with the sum of two Gaussians.
\label{vel.resid}}
\end{figure*}

\begin{figure}
\epsscale{0.90}
\plotone{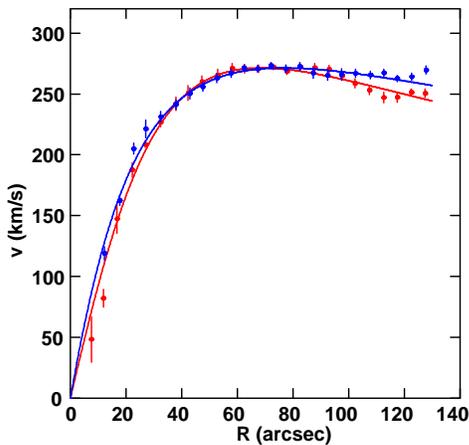}
\caption{Brandt rotation curve fits to major-axis quadrants
of the 
first moment image of NGC 3145 with the receding major-axis
quadrant in red and the approaching major-axis
quadrant in blue.
\label{rot.curve}}
\end{figure}

\begin{deluxetable*}{lcccc}
\tabletypesize{\scriptsize}
%\tablenum(5)
\tablecaption{Velocity Parameters from fitting the data with GAL\label{galfit}}
\tablewidth{0pt}
\tablehead{
   \colhead{Quadrant} & \colhead{$i$}  & \colhead{$v_{\rm max}$}
  &  \colhead{$R_{\rm max}$} &  \colhead{$n_{\rm B}$} \\
  &  \colhead{($\arcdeg$)}  &  \colhead{(\kms)} & \colhead{($\arcsec$)}\\
}
\startdata
whole disk           & $46.1 \pm 0.2$  &  $286.8 \pm 0.9$  &  $74.6 \pm 0.5$
   &  $1.44 \pm 0.02$ \\
NNE Major Axis   &  $50.6 \pm 0.3$  &  $271.4 \pm 1.0$  &  $76.5 \pm 0.9$
   &  $1.28 \pm 0.03$ \\
SSW Major Axis   &  $50.1 \pm 0.3$  &  $271.4 \pm 1.1$  &  $68.6 \pm 0.5$
   &  $1.84 \pm 0.03$ \\
\enddata
\end{deluxetable*}

On the other hand,
the velocity-residual field exhibits a curious asymmetry 
on and near the minor axis. In Figure\,\ref{vel.resid} , the residual velocity
image (observed minus model) is 
displayed in greyscale and contours,  and the locations of the kinematic 
center, apex $a$, and apex $b$ are denoted by the plus sign, the five-pointed
star, and the small triangle symbols, resp. Lines are drawn in this figure to divide 
the galaxy into four quadrants: these are centered on the receding major axis 
in the SSW, the near-side (relative to us) minor axis in the WNW, the 
approaching major axis in the NNE, and the far-side minor axis in the
ESE.  On the near-side minor axis, the residual velocities are receding from
us, whereas on the far-side minor axis, the residual velocities are 
approaching us. The minor axis is where radial or $z$-motions (perpendicular
to the disk)  would be most apparent. The observed asymmetry would be 
consistent with radial inflow or with $z$-motions.
 The asymmetry region overlaps part of the ring-like distribution of
\sixcm\  emission, so if the motions producing the asymmetry were
radial, this would argue against an expanding pseudo-ring because it would have
to be a contracting pseudo-ring instead.
We conclude below that
the more likely explanation is $z$-motions.
The residual velocities have somewhat greater magnitude on the WNW 
minor axis (25 to 62 \kms) than on the ESE minor axis  ($-19$ to $-27$ \kms).
At the location of the {\it triangle}, the residual velocities are small $\simeq 4$
\kms.

With the values of the kinematic center and systemic velocity fixed as above,
we used GAL to fit separate Brandt rotation curves to the 
receding and approaching major-axis quadrants.
 The  results are listed in
Table\,\ref{galfit} and displayed  in
Figure\,\ref{rot.curve}. Brandt curves are good fits to the major axis quadrants,
and for values of the face-on radius $R \geq 30''$, 
they yield velocity differences of less than 5\% between the approaching
and the receding sides. The solution for the major axis quadrants 
gives a value for the inclination $i = 50.3\degr \pm 0.3\degr$, 
(a little smaller than the
isophotal value $i = 55\degr$ from the {\it 2-MASS} isophotes and 
the $N$(HI) outer isophotes) and a maximum velocity of the rotation curve
$v_{\rm max} = 271 \pm 1$ \kms. This maximum occurs at $R_{\rm max}$ =
$68.6''$ = 18 kpc on the receding side and $76.5''$ 
= 20 kpc on the approaching side.

\begin{figure*}
\epsscale{1.02}
\plotone{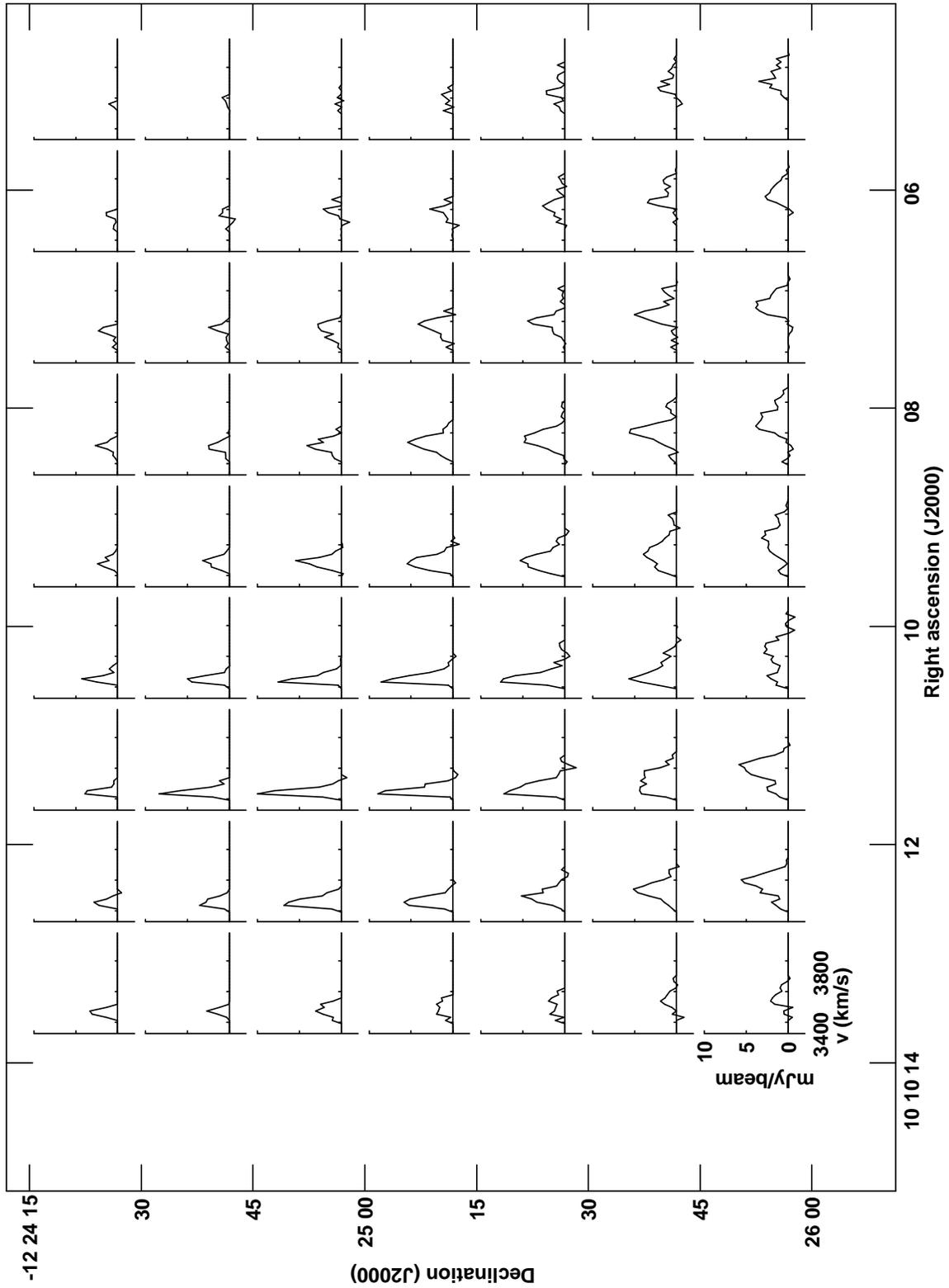}
\caption{\HI\ line-profiles spaced $15''$ apart from Cube 1
(with $\sigma_{\rm rms}$ = 0.50 \mJybeam\ and 
21.2 \kms\ resolution) for the
entire northern half of NGC 3145.
The line-profiles on
and near the major axis consist of a large  amplitude peak 
with a skewed wing towards higher velocities.
\label{profiles.north}}
\end{figure*}

\begin{figure*}
\epsscale{0.88}
\plotone{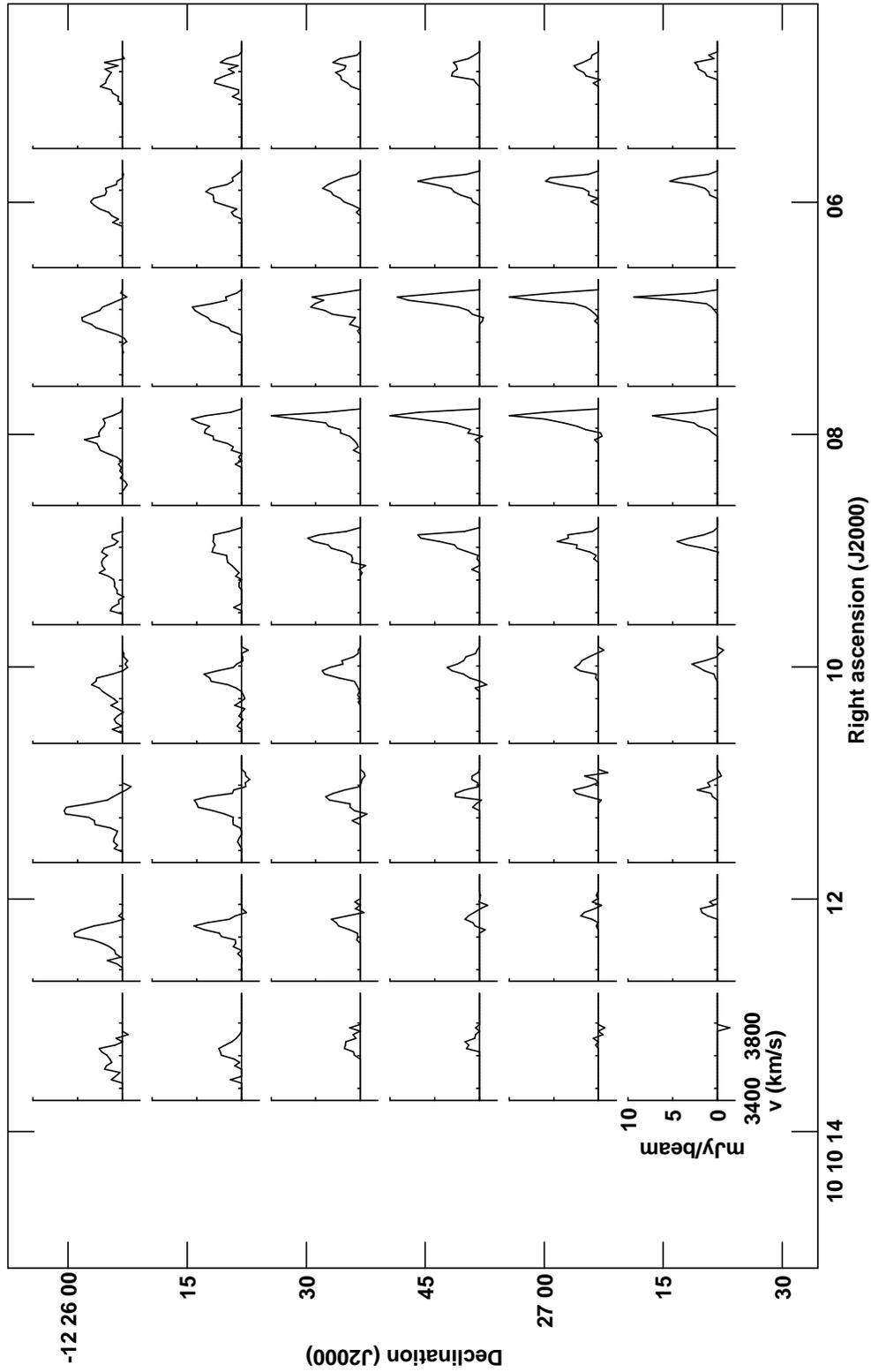}
\caption{\HI\ line-profiles spaced $15''$ apart from Cube 1 for 
the entire southern half of NGC 3145. The rms noise = 0.50 \mJybeam.
The line-profiles on
 and near the major axis consist of a large amplitude peak
 with a skewed wing towards lower velocities.
\label{profiles.south}}
\end{figure*}

To provide more information about noncircular motions, 
Figure\,\ref{profiles.north} displays \HI\ line profiles, spaced $15''$ apart
(i.e, every third spatial pixel) from Cube 1 for the entire northern half of  NGC 3145.
Figure\,\ref{profiles.south} does the same for the entire southern half of the galaxy.
At distances $\gtrsim 35''$ from the nucleus, the line profiles on and near the 
approaching major axis (NNE side) consist of a large amplitude, narrow peak 
with an asymmetric, extended, high velocity wing, and the line profiles
on and near the receding major axis (SSW) side consist of a large amplitude,
narrow peak with an asymmetric, extended, low velocity wing.
 Since observed motions on a major axis cannot contain a contribution 
from radial inflow or radial outflow,  the possible interpretations of the skewed
wings in these profiles are out-of-plane gas in a thick disk with a slower 
rotation speed than the thin disk or
$z$-motions of gas moving away from us (relative to the 
disk) on the approaching major axis and towards us 
(relative to the disk) on the receding major axis. As we find kinematic
oddities on the minor axis as well as on the major  axis, the simplest 
explanation for both is $z$-motions. Given the optical oddities 
that suggest out-of-plane arms in this galaxy, it is not surprising to find motions
perpendicular to the disk. 
Taken together with the asymmetry in the residual velocities on and near
the minor axis (see Figure\,\ref{vel.resid}), it seems that relative to the disk,
there is gas with $z$-motions away from us on and near the NNE major
 axis and near the WNW minor axis 
and $z$-motions towards us on and near the SSW major axis and near the 
ESE minor axis. This anti-symmetry of
the $z$-motions suggests  the disk is undergoing warping as a result of a close 
passage by a companion. We detect these warping motions out to a radius
of at least $105''$.

For the line-profiles near the major axis that consist of just a
major peak  plus a skewed tail, we fit the sum of two Gaussians 
to each profile in the two boxes marked in Figure\,\ref{vel.resid}. These 
$35'' \times  35''$ boxes
are centered approximately at $R_{\rm max}$ on the major axis, and
we shall refer to them as the $R_{\rm max}$ boxes
We assume that the large-amplitude peak in the profile is from gas 
in the disk
and that the skewed tail is from gas moving perpendicular to the disk.
The goals are to determine some information about the \HI\ gas
involved in the $z$-motions and to determine how much the rotation curve 
derived from the \HI\ velocity field at these positions is affected by the
$z$-motions.  Let the subscript 1 refer to the  parameters from
the Gaussian fit at the large-amplitude peak in each line profile 
and subscript 2
refer to the parameters from the Gaussian fit  to the skewed tail. 
Table\,\ref{Rmax.box} 
lists the mean values from the two-Gaussian fits to the 
following parameters for the $R_{\rm max}$ boxes: (a)
the difference  $v_1 - v_2$ in the line-of-sight central velocities
 of the Gaussians,
 (b) the difference $v - v_1$ between the velocity $v$ of the velocity
field in the first moment image and the central velocity $v_1$ 
of the large-amplitude peak, (c) the FWHM widths of the Gaussians, 
and (d) the ratio $N({\rm HI})_2/N({\rm HI})_1$ of the column densities.
Locations with bad solutions or large uncertainties
are omitted. The uncertainty attached to 
each mean is the mean of the Gaussian uncertainties. Also listed
are the values of the standard deviation $s$ of the sample. 

\begin{figure*}
\epsscale{1.0}
\plotone{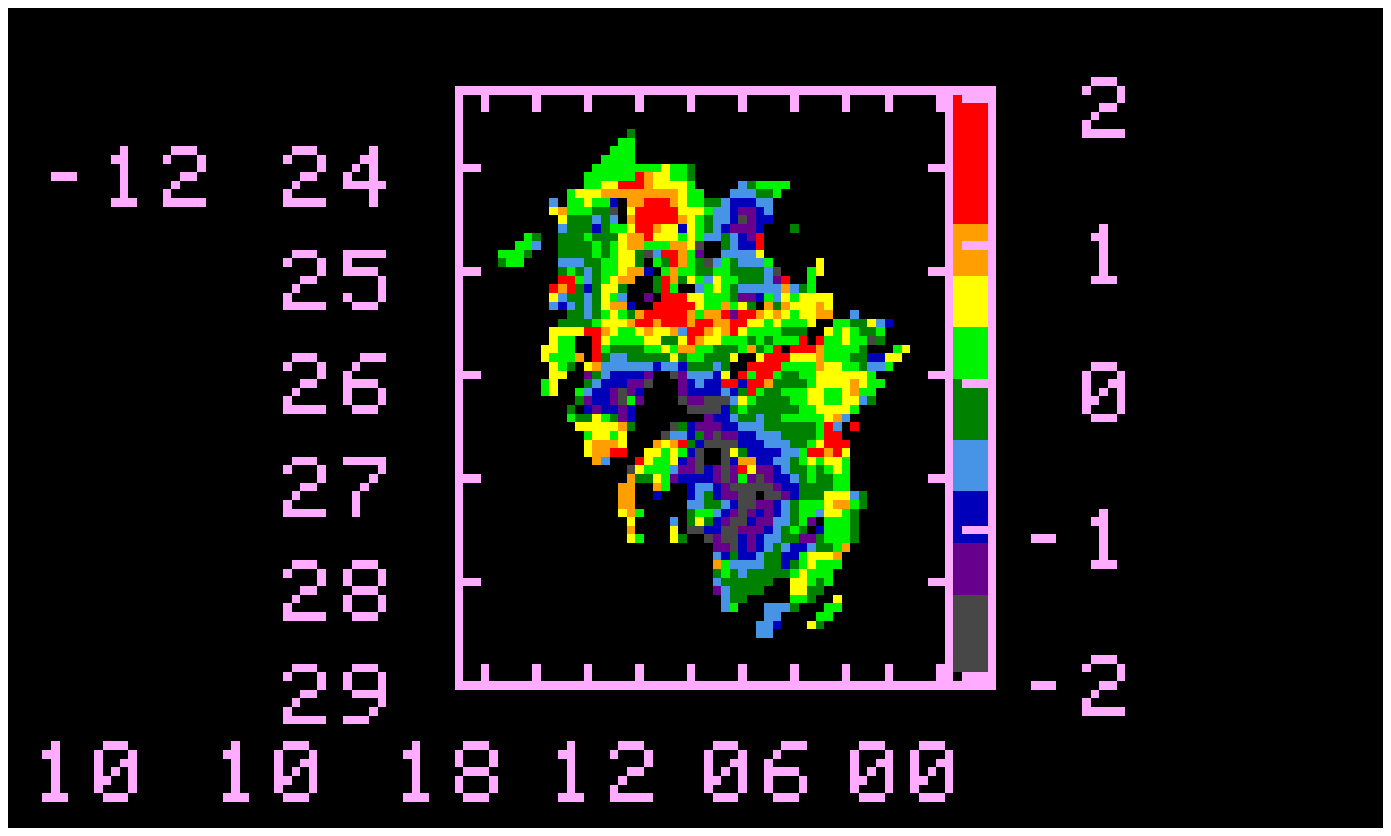}
\caption {Display of the Fisher-Pearson coefficient of skewness 
of the line-profiles in NGC 3145.
Regions in green have negligible skewness of the line-profiles. 
The rest have moderate to high skewness. {\it Most of the half 
of the galaxy from PA = $60\degr$ to PA = $240\degr$ has negative 
skewness  of the line profiles, i.e., gas moving towards us 
relative to the disk. The other half of the galaxy has large 
areas with positive skewness, i.e., gas moving away from us
relative to the disk.}
\label{skewness}}
\end{figure*}

\begin{deluxetable*}{lccccc}
\tabletypesize{\scriptsize}
%\tablenum(6)
\tablecaption{Two-Gaussian Fits to Line-Profiles in $R_{\rm max}$ Boxes\label{Rmax.box}}
\tablewidth{0pt}
\tablehead{
  \colhead{Parameter} & \colhead{Receding Side} &&& \colhead{Approaching Side}&\\
    & \colhead{mean} & \colhead{$s$\tablenotemark{a}} 
    && \colhead{mean} & \colhead{$s$}\\
} 
\startdata
$v - v_1$ (\kms)      & $-15\pm 1$  & $\pm 4$ &&  $12 \pm 1$  & $\pm 6$\\
$v_1 - v_2$ (\kms)  &  $47 \pm 7$  & $\pm 11$ && $-37 \pm 3$ & $\pm 7$\\
$(FWHM)_1$ (\kms) &  $41 \pm 5$  &  $\pm 6$ && $31 \pm 4$ & $\pm 3$\\
$(FWHM)_2$ (\kms) &  $67 \pm 12$ & $\pm 10 $ && $50 \pm 9$ & $\pm 10$\\
$N({\rm HI})_2/N({\rm HI})_1$  &  $0.4 \pm 0.1$ & $\pm 0.2$ && 
           $0.9 \pm 0.2$ & $\pm 1.0$\\
\enddata
\tablenotetext{a} {$s$ is the standard deviation of the sample}
\end{deluxetable*}

On the receding major-axis, the mean value of $v - v_1$  for the
$R_{\rm max}$ box is $-15 \pm 1$ \kms\ with $s$ = 4 \kms,
and for positions closest to the major axis the mean is
 $-11$ \kms\ with $s$ = 2 \kms. On the
approaching major axis, the mean value of $v - v_1$
for the $R_{\rm max}$ box is $12 \pm 1$ \kms\
with $s$ =  6 \kms, and for positions closest to the major axis the
mean is 8 \kms\ with $s = 3$ \kms.  Therefore
the  correction for this effect on 
$v_{\rm max} \sin i $ amounts to only about 5\%,
 which is negligible.

For a spherically symmetric mass distribution, the dynamical mass 
$M_{\rm dyn}(R)$ within face-on 
radius $R$ is $M_{\rm dyn}(R)$ = $R (v_{\rm rot})^2/G$
= $2.33 \times 10^5 R (v_{\rm rot})^2$  \msun\ for 
$R$ in kpc and circular velocity $v_{\rm rot}$ in \kms.
 We fit the velocity data in the major axis quadrants 
out to $R$ =$127''$ = 33.3 kpc, which is (approximately)
the  extent of 
significant major-axis emission in the $N$(HI) image. At
this radius, the fitted 
curves in Figure\,\ref{rot.curve} have an average
$v_{\rm rot}$ = $252 \pm 6$  \kms, which 
 gives a dynamical mass $M_{\rm dyn}$ =
$5.0 \times 10^{11}$ \msun. Since $M$(HI) =
$1.4 \times 10^{10}$ \msun, \HI\ accounts for only 3\% of
the dynamical mass out to this radius. The radius of $127''$
that we use for $M_{\rm dyn}$ is greater than the Holmberg
radius of $103''$ used by \citet{faber79} for $M_{\rm dyn}$. 
At R = $103''$ our data have an average $v_{\rm rot}$ 
  of $263 \pm 3$ \kms,  which is only 5\%
greater than the value of 250 \kms\ used by \citet{faber79}.
Thus aside from the difference in the adopted distance 
of NGC 3145, our value for $M_{\rm dyn}$ out to the
Holmberg radius is consistent with theirs.

The magnitude of the difference $|v_1 -v_2|$ in the 
line-of-sight
central velocities in the $R_{\rm max}$ box on the receding side
$47 \pm 7$ \kms\ 
is marginally consistent, within the uncertainties, with its 
magnitude $37 \pm 3$ \kms\ in the $R_{\rm max}$ box
on the approaching side.
 This is compatible with the suggestion that the skewed tail 
in these line profiles
was  produced by the same mechanism on both sides of the
galaxy.

On both sides, the FWHM width of the Gaussian fit to the line profile 
 is about 20 \kms\ broader for the skewed tail than for
large amplitude peak. The larger velocity width for the skewed tail
could result from a velocity gradient along the flow .

For the ratio $N({\rm HI})_2/N({\rm HI})_1$ of the column 
densities from the Gaussian fits, we restrict to positions where the
propagated uncertainty in this ratio is less than 50\%; this eliminates 
about half of the positions in the above boxes. Then for the 
$R_{\rm max}$ box on the receding side,
the mean value of $N({\rm HI)}_2/N({\rm HI})_1$ = $0.4 \pm 0.1$
with standard deviation $s$ of the sample = 0.2 and range 
 = 0.2 to 0.9.  For the $R_{\rm max}$ box on the approaching
side,  the mean value of this ratio
= $0.9 \pm 0.2$ with $s$ = 1.0 and range = 0.20 to 3.5.
We conclude that the amount
of gas involved in the $z$-motions is not insignificant but varies
considerably with position.
 
Another way of analyzing the velocities in NGC 3145
 is to look at the image of the skewness of the line-profiles
in Figure\,\ref{skewness}. The parameter displayed is the
dimensionless Fisher-Pearson coefficient of skewness 

\begin{displaymath}
    g_1 = \frac{m_3}{(\sigma_{\rm v})^3}
\end{displaymath}
where $m_3$ is the third moment of the velocity data. For a sample
 of size $n$, $g_1$ needs to be multiplied by a correction factor 
\begin{displaymath}
   \frac{[n(n-1)]^{1/2}}{n-2}.
\end{displaymath} 

For Cube 1, this correction factor varies from 1.06 near the center 
to 1.2 in the outskirts of NGC 3145. In Figure\,\ref{skewness}, regions
in green have negligible skewness of the line-profiles, i.e., 
$|g_1| < 0.5$, and the rest of the galaxy has moderate to high
skewness of the line-profiles.
This image demonstrates that (1)  most of the half of the galaxy
from position angle PA = $60\degr$ to  PA = $240\degr$
  has negative skewness of the line-profiles, 
i.e., gas moving towards us relative to the disk.
(2) In the other half of the galaxy there are
large areas of positive skewness, i.e., gas moving away
from us relative to the disk.

\subsubsection{Small Scale Anomalies}

We study the various oddities marked in Figure\,\ref{B.Features}
by inspecting their \HI\ line-profiles. For NGC 3145, we find
the line-profiles to be more revealing than position-velocity 
diagrams are.

\begin{deluxetable}{lcc}
\tabletypesize{\scriptsize}
%\tablenum(7)
\tablecaption{Corrected Velocity Dispersions\label{corr.disp}}
\tablewidth{0pt}
\tablehead{
  \colhead{Location} & \colhead{$\sigma_{\rm v}$}  & 
      \colhead{ Corrected $\sigma_{\rm v}$}\\
& \colhead{(\kms)}  & \colhead{(\kms)}
}
\startdata
apex $a$                              & 94  &  68\\
opp. apex $a$                      & 60  &  40\\
\\
apex $b$                              & 50  &  36\\
opp. apex $b$                      & 41  &  38\\
 \\
apex $c$                              & 50  &  36\\
opp. apex $c$                      & 57  &  50\\
 \\
{\it triangle} (mean)              & 72   &  49\\
opp. {\it triangle} (mean)      & 50  &   37\\
 \\
Feature $f$                          &111 &   83\\
opp. Feature $f$                  & 98  &   85\\
 \\
complex dust-loops (mean)  & 93  &  79\\
Feature $d$ (mean)               &  71  &  59\\
western antenna (mean)        &  54  &  46\\
\enddata
\end{deluxetable}

For features of interest, Table\,\ref{corr.disp} lists values 
 of  \sigmav\ of the \HI\ gas before and
after correction for the mean velocity gradient across the PSF.
These are from Cube 1, the cube with 21  \kms\ velocity
resolution. The correction for these features
decreases the velocity dispersion by 15\% to 32\%. 
For comparison, this Table also contains the values of \sigmav\
at the diametrically-opposite locations of some of the
features. These are listed as ``opp. apex $a$,'' ``opp. apex $b$,'' etc.
Since the line-profiles are not corrected for the velocity gradient,
one does well to keep in mind the size of the correction to \sigmav\
when viewing the line-profiles since a large velocity gradient 
plus finite spatial resolution artificially broaden a 
line-profile. Our method of correcting the velocity dispersion is 
less accurate in regions, such as near the galaxy center, where 
the velocity gradient changes significantly across the PSF. 
This inaccuracy contributes to the very high values of
the corrected \sigmav\ in part of the region of complex 
dust loops and at Feature $f$ (see velocity contours in
Figure\,\ref{N(HI).N3145}).

All of the corrected values of \sigmav\ for the features in 
Table\,\ref{corr.disp}  
(36 \kms\ to 68 \kms, if we exclude Feature $f$ and
the region of complex dust  loops) are
large compared to the values of 6 - 13 \kms\  found by 
\citet{kamphuis93} for the \HI\ gas
in undisturbed spirals,  but are comparable to the high values
of \sigmav\ measured in some galaxy pairs undergoing close
encounters \citep{elmegreen95, kaufman97, kaufman99}.

\begin{figure*}
\epsscale{0.874}
\plotone{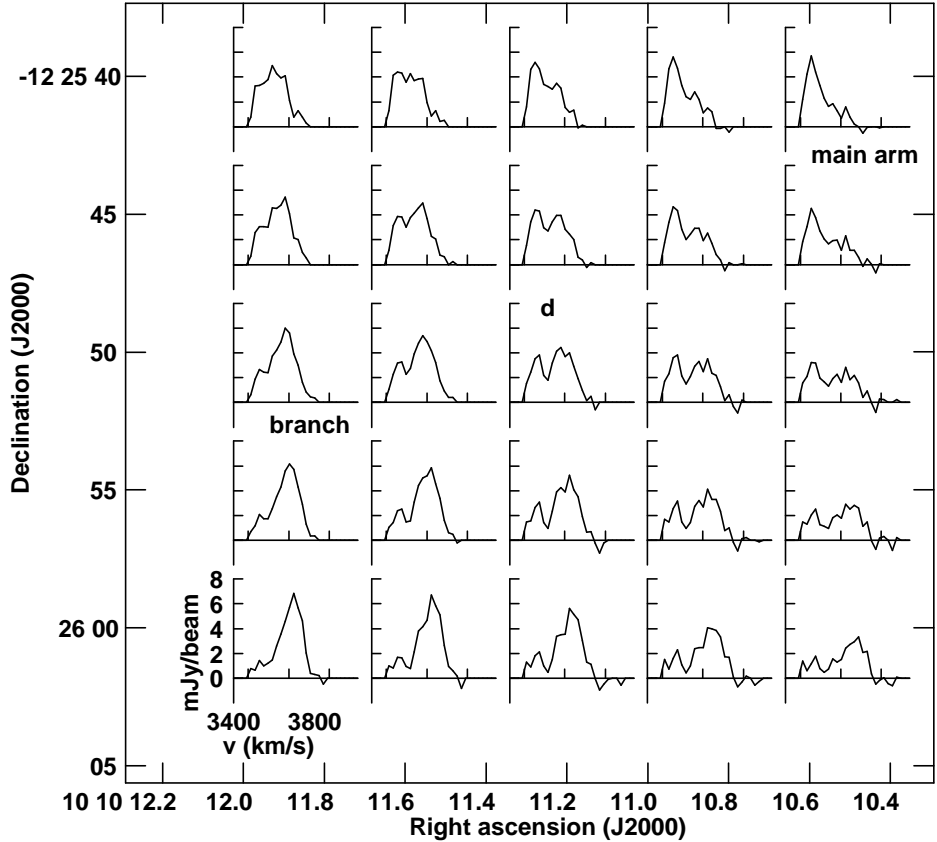}
\caption{\HI\ line-profiles of the main spiral arm and the
{\it branch} on the northeastern side of NGC 3145 from
Cube 1, spaced $5''$ apart. The rms noise is 0.5 \mJybeam.
The panel labelled ``d'' is
where the {\it branch} appears to intersect the
main spiral arm. The {\it branch}
is the peak at 3623 \kms, and the main spiral arm is
the peak at 3475 \kms\ in the line-profile. The 
$\sim 150$  \kms\ difference in velocity between the
{\it branch} and the main spiral arm where in projection 
they appear to intersect 
 implies that {\it the branch is a tidal arm moving away
from us.}
\label{branch}}
\end{figure*}

The clearest situation is Feature $d$ in Figure\,\ref{B.Features} .
Figure\,\ref{branch} displays \HI\ line-profiles, spaced $5''$ 
apart, for Feature $d$ and its surroundings. 
The panel labelled
``d" is where the branch and the main spiral arm 
appear to intersect,
and the panel just below it is where the {\it branch} departs
from the spiral arm.
The four panels in the upper right corner of  Figure\,\ref{branch} 
(i.e., the intersection of rows 4 and 5, with 
columns 4 and 5, where we number rows and columns from the 
bottom left corner of the figure) 
are located along the main spiral arm. 
The greatest amplitude peak in their 
line-profiles is at 3454 \kms. The six panels in the bottom left 
corner of the figure  (the intersection of rows 1, 2, and 3 with 
columns 1 and 2)  are located along the {\it branch}. The greatest amplitude 
peak in their line-profiles is at $\sim 3620$ \kms.  The line-profiles
at the panel where the {\it branch} and the main arm 
appear to intersect
and at the panel where the {\it branch} departs from the main 
arm  have two main peaks: one at 3623 \kms\
(i.e., the {\it branch}) and the other at 3475 \kms (i.e., the 
main spiral arm).
Although optically feature $d$ looks like a branch of 
the spiral arm, the $\sim 150$ \kms\ difference in line-of-sight
velocity between the {\it branch} and the main spiral arm 
where  in projection they appear to intersect suggests that
the {\it branch} is an out-of-plane feature moving away from us 
relative to the disk, and there may be large streaming motions along 
it. Thus  {\it the  branch is a tidal arm.}
To produce the observed structure, the perturbation needed
an azimuthal component to draw out the tidal arm and a perpendicular
component to make it extra-planar. The 
apparently major \HI\ concentration at this location cannot be a single entity; 
instead one part is associated with the main spiral arm and another part 
 with the {\it branch}.  As noted in Section 4.1 (see
Figure\,\ref{N(HI).N3145}),  the brighter \HI\ here is elongated
to the south, not along the main spiral arm but along
the {\it branch} or a bit east of it, and the velocity field contours 
have wiggles along the {\it branch}.

\begin{figure}
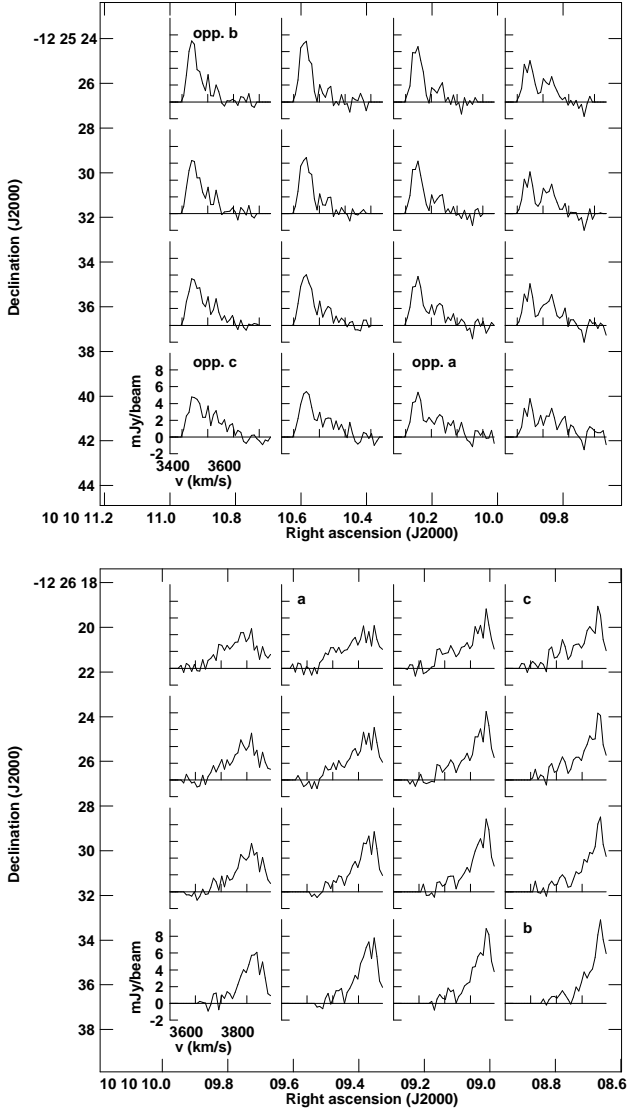

\epsscale{1.17}
\plotone{fig21bot.eps}
\plotone {fig21.eps}
\caption{Bottom: \HI\ line-profiles from the cube with 10.6 \kms\
resolution, spaced $5''$ apart, for the {\it triangle}.
The rms noise is 0.74 \mJybeam.
The panels at apices $a$, $b$, and
$c$ of the {\it triangle} are labelled ``a,'' ``b,'' and ``c,'' resp. 
 Top: \HI\ line-profiles for the diametrically-opposite region.
The line-profiles labelled ``opp. a", ``opp. b'', and ``opp. c''
are diametrically-opposite apices $a$, $b$, and $c$, resp.
Only apex $a$ differs significantly in shape and width of the
line-profile from that of its diametrically-opposite counterpart.
\label{profiles.10kms}}
\end{figure}

\citet{struck90} comments that the optical {\it triangle} 
on the southern side of NGC 3145 looks like a swallowtail caustic.
A swallowtail caustic is 
composed of five ``intersecting'' star streams.  In his study of
two-dimensional caustics with collisionless particles,
 he finds that such features may result
from a slightly off-center direct collision with a smaller galaxy.

Our \HI\ data do not have sufficient resolution to see individual gas
streams in the {\it triangle}, and we note that once gas is included, a
three-dimensional model would be needed. Instead, as a test we 
compare  the \HI\ line-profiles in the {\it triangle} with those in the
diametrically-opposite region on the northern side of NGC 3145.  
Also for the features of interest, we compare the values of the corrected velocity
dispersion in Table\,\ref{corr.disp} with those at the diametrically-opposite locations.
In Figure\,\ref{profiles.10kms}, the bottom panel displays 
\HI\ line-profiles, spaced $5''$  apart, for the {\it triangle}
from Cube 3, which has  a velocity resolution of 10.6 \kms\ 
 and a spatial resolution of
 $27.3'' \times 16.6''$ (HPBW), 
BPA = $- 29\degr$. The top panel displays the 
diametrically-opposite region, with line-profiles labelled 
``opp. a,''``opp b,'' and ``opp. c.''
Only apex $a$ shows a significant difference when comparing 
its line-profile and corrected \sigmav\ with those of the
diametrically-opposite location. 
Apex $a$ is where the inner spiral arm and Sandage's peculiar
arm intersect in projection. Thus, our data cannot confirm the swallowtail
interpretation of the apparent {\it triangle}.

If a slightly off-center direct collision were responsible for the 
anomalous features in NGC 3145, then to
produce the observed warping motions over large spatial scales found in
Section 4.2.1, the orbital tilt angle 
 would have to be significantly less than $90\degr$
relative to the disk of NGC 3145.  Our failure to find a strong expanding 
ring is consistent with an encounter that is not close to perpendicular.

The line-profile at apex $b$ (see Figure\,\ref{profiles.10kms})
consists of a large amplitude,  narrow peak plus an asymmetric,
extended wing to lower velocities.
Fitting the sum of two Gaussians 
to the line-profile at apex $b$  in Cube 1
gives (a) for  the large amplitude peak, a central velocity $v_1$
= $3872 \pm 1$ \kms\ and $(FWHM)_1$= $42 \pm 3$ \kms, 
and (b) for the skewed wing, a central velocity $v_2$ =
$3818 \pm 11$ \kms\ and $(FWHM)_2$ = $100 \pm 17$ \kms.
Thus $v_1 - v_2$ =$54 \pm 11$ \kms.  These values for 
the line-of-sight velocity difference $v_1 - v_2$
and $(FWHM)_1$ are similar to the mean values listed in  
Table\,\ref{Rmax.box} for the warping motions in the receding-side 
$R_{\rm max}$ box (which is southwest of apex $b$ and the 
western antenna).
The value of $(FWHM)_2$  for apex $b$ is somewhat greater
 than the mean for this
$R_{\rm max}$ box. Unlike apex $b$,
the $R_{\rm max}$ box does not contain an ``X''-feature (where
two arms appear to cross in projection).
We note that two arms having the same line-of-sight velocity where they
produce an ``X''-feature could, nevertheless, be in different planes.

\begin{figure}
\epsscale{1.1}
\plotone{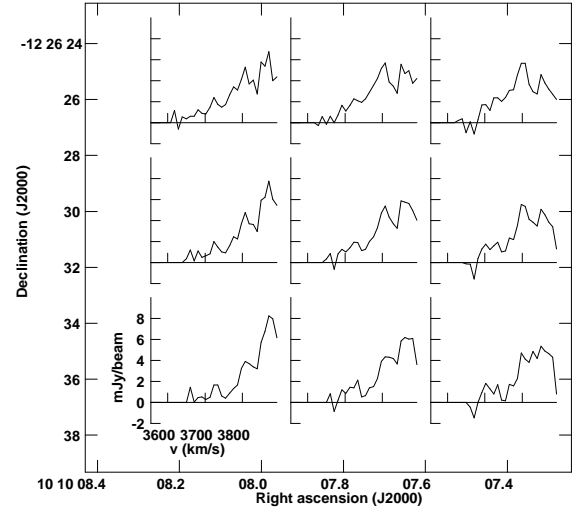}
\caption{\HI\ line profiles of the western-antenna  
continuation of Sandage's peculiar arm. These are from
the cube with 10.6 \kms\ resolution, spaced $5''$
apart. The rms noise is 0.74 \mJybeam.
All but one of these line-profiles have two clearly distinct 
peaks separated by $56 \pm 8$ \kms. We attribute one
of these peaks to Sandage's peculiar arm and the other 
to the disk.
\label{west.antenna}}
\end{figure}

The western antenna is a continuation of Sandage's 
peculiar arm. Figure\,\ref{west.antenna} displays
 western antenna line-profiles 
of feature $e$ in Figure\,\ref{B.Features}. These are 
 for the smaller of the two boxes drawn on the {\it HST}
image in that Figure.
 The line-profiles are from Cube 3 (the
cube with 10.6 \kms\ resolution), spaced $5''$ apart and
do not resemble those at the diametrically-opposite location.
Eight of the nine line-profiles in this figure exhibit two main, 
clearly distinct peaks of comparable amplitude. The mean 
separation in velocity between the two peaks is 56 \kms\ with a
standard deviation $s$ of the sample = 8 \kms. This is essentially the
same as the value of  $v_1 - v_2$ = $54 \pm 11$ \kms\ at
apex $b$.
We attribute one of these peaks to Sandage's peculiar arm
and the other to the disk. 
Like the {\it branch} on the northeastern side of NGC 3145,
Sandage's peculiar arm appears to be  a tidal arm, but 
 differs by only $56 $ \kms\ in line-of-sight velocity
from that of the disk, whereas the {\it branch} differs by
$\sim 150$ \kms\ from that of the disk.

 In Section 3.2, we found that the portion of Sandage's 
peculiar arm from Feature $f$ northwards appears to be
 in the disk but once this arm reaches the eastern side
of the {\it triangle}, it is no longer in the disk.  If  
Sandage's peculiar arm is a tidal arm coming from the central
part of the galaxy, it is not surprising to find a shock front 
along its initial part until it emerges from the disk.

In summary, we find two types of extra-planar motions in NGC 3145:
(i) an anti-symmetric,  global warping of the disk
 and (ii) extra-planar tidal arms.  Feature $d$ is quite distinct from
the warping motions on the northern side of the galaxy
in terms of the size of the line-of-sight velocity difference between 
components of the line-profile (150 \kms\ versus  37 \kms)
and the shape of the line-profile.  In contrast to this,
the line-of-sight  velocity difference between
Sandage's peculiar arm and the disk is similar in size to that 
of the warping motions.  For these two extra-planar arms, the
observed difference in line-of-sight velocity  between the arm and
the disk could be a combination of $z$-motions and streaming
along the arm. We do not have sufficient information to determine
the relative contributions of these two types of motions.

\section{\HI\ Properties of NGC 3143 and PGC 029578}

Basic \HI\ properties of the two companions, NGC 3143 and 
PGC 029578,  are listed in
Table\,\ref{HIproperties}.  Both companions are on the receding side
of NGC 3145 (see Figure\,\ref{N(HI).N3145}) and have 
values of $v_{\rm sys}$ less than that of NGC 3145. 
There are no other galaxies near NGC 3145 that 
have published optical redshifts close to that of NGC 3145 or have
 \HI\ detections.

The SB(s)b galaxy NGC 3143 at $8.97\arcmin$ (= 141 kpc) south of
NGC 3145 is the nearer to it of the two companions and has the 
smaller diameter.
It is bluer in $B - V$ than NGC 3145 (see Table\,\ref{basic.optical}), and 
optically its spiral arms are brighter than the spiral arms of NGC 3145.
In the top panel of Figure\,\ref{N3143.HI}, $N$(HI) contours of NGC 3143 
are overlaid on the $B$ image from the Burrell-Schmidt 
telescope. The \HI\ column densities in NGC 3143 are 
significantly lower  than in NGC 3145 or in PGC 029578
(see Table\,\ref{HIproperties} and  Figure\,\ref{HI.triplet}). 

In $K_s$-band, the luminosity of NGC 3143 is 0.073 times the
luminosity of NGC 3145. We adopt this as the stellar mass ratio
of  NGC 3143 to NGC 3145. If we assume that the total mass 
ratio of NGC 3143 to NGC 3145 is the same as the stellar mass ratio, 
then since the dynamical mass of 
NGC 3145 is $M_{\rm dyn}$ =$5.0 \times 10^{11}$ \msun\ out
to  $1.3 \times$ the optical radius, 
 the estimated mass of NGC 3143 is $\sim 3.7 \times 10^{10}$ \msun.

The following
comparisons indicate that NGC 3143 is somewhat deficient in \HI.
The \HI\ mass $M$(HI) of NGC 3143 is $6.5 \times 10^8$ \msun,
which is only 4.5\% of the \HI\ mass of NGC 3145.
 Thus the ratio of $M$(HI) to stellar mass is a factor of 1.6 greater 
 in NGC 3145 than in NGC 3143.  
In NGC 3143, the ratio $M$(HI)/$L_{\rm B}$ =
0.085 \msun $L_{\sun}^{-1}$. This value
lies in the bottom 
25\% of  Sab, Sb spirals in \citet{roberts94}
and is half of its value in NGC 3145.
 This low value of $M$(HI)/$L_{\rm B}$ could be the result of more
active star formation in NGC 3143 as indicated by its bluer 
$B - V$ color or of loss of some \HI\ by NGC 3143 in an
encounter. The amount of molecular gas in these galaxies has 
not been measured.

\begin{figure*}
\epsscale{0.7}
\plotone{fig24a.eps}
\plotone{fig24b.eps}
\plotone{fig24c.eps}
\caption{Top: $N$(HI) contours of  NGC 3143
overlaid on a $B$ image in greyscale. Contour levels are
at 100, 200, and 300 \Jybeamms,
 where 100 \Jybeamms\
corresponds to  $N$(HI) = $2.3 \times 10^{20}$ \atc.
NGC 3143 is somewhat deficient in \HI, and its
\HI\ emission is more elongated on the 
northeastern major axis than on the southwestern major 
axis. Middle: \HI\ channel maps of NGC 3143
with contours at (3, 4, 5, 6) 
$\times$ the rms noise of  0.74 \mJybeam
(equivalent to 1.0 K).
Location of the nucleus is
marked by a cross. The western side of NGC 3143
is the receding side.
Bottom: Contours from the first moment map
overlaid on the $N$(HI) image in greyscale.
\label{N3143.HI}}
\end{figure*}

\begin{figure*}
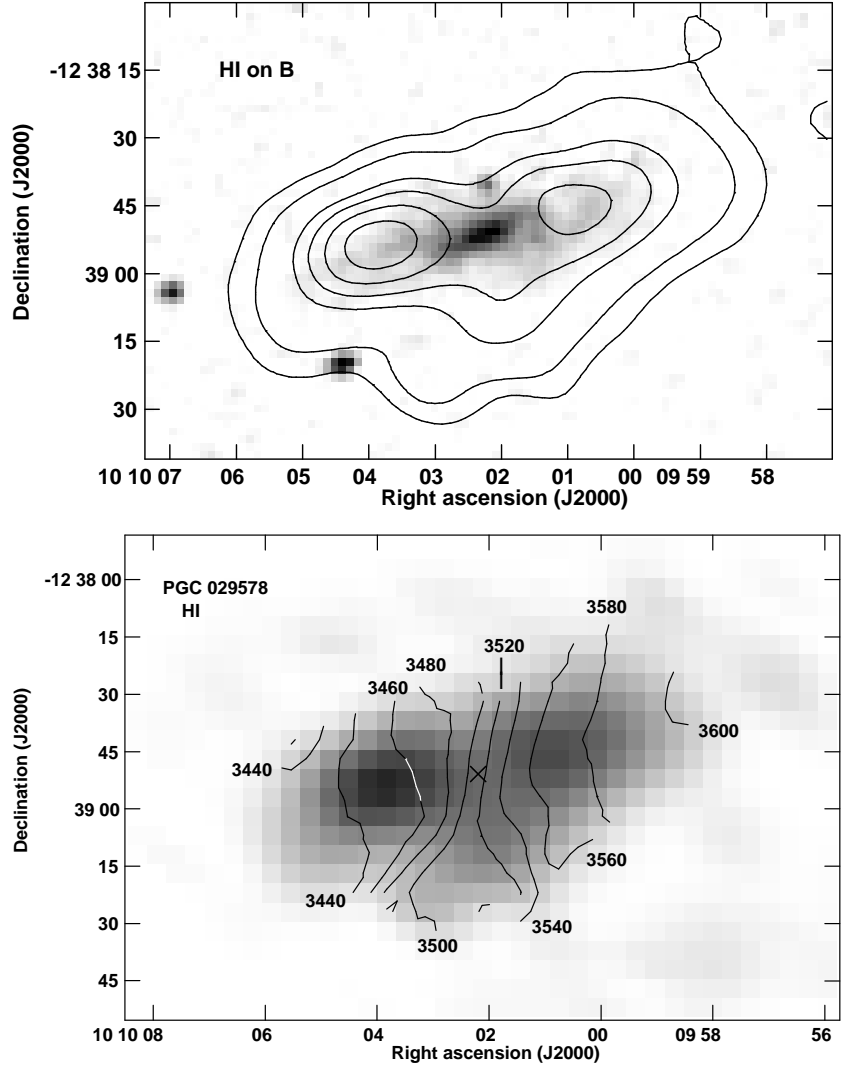

\epsscale{0.75}
\plotone{fig25a.eps}
\plotone{fig25b.eps}
\caption{Images of PGC 029578. Top: $N$(HI) contours
overlaid on a DSS image in greyscale. The contour
levels are 100, 200, 400, 500, 600, 700, 800
\Jybeamms, where 100 \Jybeamms\ corresponds
to $N$(HI) = $2.3 \times 10^{20}$ \atc. 
Bottom: Contours from the first moment image
overlaid on the $N$(HI) image in greyscale, with 
a cross symbol marking the location of the nucleus.
\label{PGC.HI}}
\end{figure*}

We estimate the star formation rate (SFR) of NGC 3143 from the 
magnitudes listed by {\it WISE}.  At 22 \micron\ NGC 3143 has 
a magnitude $m$(22 \micron) = $6.173 \pm 0.055$ mag.  
Since it is 4 mag brighter at 12 \micron\ than at 4.6 \micron,
NGC 3143 is a red source in the mid-infrared. 
To convert  $m(22\ \micron)$  to flux density $S_\nu$(22 \micron),
we use the 
zero-magnitude flux-density, the correction for red sources, and other
information from the online {\it WISE} All-Sky Release Explanatory 
Supplement.  This gives
 $S_\nu(22\ \micron)$= 26 mJy for NGC 3143.
We extrapolate to 24 \micron\ by taking $S_\nu \propto \nu^{-2}$.
Then with a distance $D$ of 54.8 Mpc  and $L(24\ \micron)$ = $\nu L_\nu$,
 the luminosity
$L(24\ \micron)$ of NGC 3143 is
$1.4 \times 10^{42}$ erg s$^{-1}$. \citet{calzetti07} find the
following relation between SFR in \msun\ yr$^{-1}$ and 
$L(24\ \micron)$ in erg s$^{-1}$,

\begin{displaymath}
 SFR = 1.27 \times 10^{-24}[L(24\ \micron)]^{0.885}.
\end{displaymath}
Applying this to NGC 3143 gives SFR = 0.25 \msun\ yr$^{-1}$.
If the molecular gas depletion time of 2.2 Gyr from \citet{leroy12}
applies to NGC 3143 as a whole, then a SFR of 
0.25 \msun\ yr$^{-1}$ would require $M$(H$_2$) = $5 \times 10^8$
\msun\, about the same as its measured \HI\ mass of $6.5 \times 10^8$
\msun.  For a small galaxy it is unusual to have such
 a large fraction of
the interstellar gas in molecular form.

{\it 2MASS} $K_s$ isophotes find that the 
isophotal major axis of NGC 3143 is at a position angle of
$225\degr$. Figure\,\ref{N3143.HI} reveals that \HI\ 
emission extends about 50\% farther on the northeastern 
major axis than on the southwestern  major
axis. At the lowest
contour level [$N$(HI) = $2.3 \times 10^{20}$ \atc] in 
Figures\,\ref{HI.triplet} and \,\ref{N3143.HI}, 
the semimajor axis of NGC 3143  is $31.5''$ on the northeastern 
side and $20.5''$ on the southwestern side, as if 
an interaction either truncated the \HI\ disk on its 
southwestern side or elongated it on its northeastern 
side. At this contour level, the \HI\ diameter of the major 
axis is $52'' \pm 2''$ for NGC 3143,
$262'' \pm 2''$ for NGC 3145,  and 
$122 \pm 2''$ for PGC 029578. The ratio of this diameter to
$D_{25}$ is $1.00 \pm 0.17$ for NGC 3143,
$1.35 \pm 0.09$ for NGC 3145,  and $1.20 \pm 0.20$
for PGC 029578. 

The middle panel of Figure\,\ref{N3143.HI} displays all the channel
maps of NGC 3143 with emission greater than 
3 $\times$ the rms
noise in Cube 3, the cube with 10.6 \kms\ velocity resolution. The
location of the nucleus is marked by a cross.  These
channel maps indicate that the receding side is on the western side
and the approaching side is on the eastern side.  

In the bottom panel of Figure\,\ref{N3143.HI}, contours from the
first moment image of NGC 3143 from Cube 3 are overlaid on 
the $N$(HI) image in greyscale.  
Since the major axis diameter is only about 3 times the HPBW of 
the \HI\ synthesized beam, we don't get much of a velocity field. 
Some estimates of the inclination $i$ from the axis ratio are 
$i$ = $35\degr$ from 
 2MASS $K_s $ isophotes and $i$ = $14\degr$ from the DSS image
\citep{zaritsky97}.  
The value of the axis ratio from the $N$(HI) isophotes depends
 on whether we use the northeastern or the southwestern
semimajor axis; with the northeastern semimajor axis
 we obtain  $i = 35\degr$, whereas with the southwestern
semimajor axis we get $i \leq 25\degr$.
At the center
of NGC 3143, the mean \HI\ velocity =  $3530 \pm 5$ \kms, which is 
consistent with the optical value of  $v_{\rm sys}$ =3536 \kms\ from 
 \citet{schweizer87}
and marginally consistent with the optical value
$3506 \pm 20$ \kms\ at the nucleus measured by \citet{zaritsky97}.
Our \HI\ value of $v_{\rm sys}$ for NGC 3143 is
$125 \pm 5$ \kms\ below the value of  
$v_{\rm sys}$ = $3655.9 \pm 0.2$ \kms\ for NGC 3145.
Therefore  an 
encounter between these two galaxies could have had a 
significant component perpendicular to the disk of NGC 3145.

The value of $v_{sys}$ = $3652 \pm 5$ \kms\ listed on NED 
for NGC 3143 from \citet{paturel03} suffers from confusion
 with NGC 3145. 

The other companion, the  Sdm galaxy PGC 029578, is 
$12.97\arcmin$ (= 204 kpc)
south of NGC 3145. The  \HI\ mass of PGC 029578, 
$3.3 \times 10^9$ \msun, is 23\% of $M$(HI) of NGC 3145
and five times the \HI\ mass of NGC 3143.

\begin{figure}
\epsscale{0.75}
\plotone{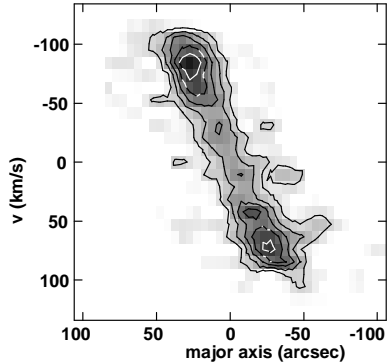}
\caption{Position-velocity diagram of PGC 009578
along its kinematic major axis as a display of its rotation
curve. Contours at (2, 4, 6, 8, 10) $\times$ rms noise 
of 0.74 \mJybeam, equivalent to 1.0 K.
\label{PGC.rotcurve}}
\end{figure}

The top panel in Figure\,\ref{PGC.HI} displays $N$(HI) contours 
of this galaxy overlaid on the DSS image in greyscale, and the 
bottom panel displays the velocity field contours from 
Cube 1 overlaid on the $N$(HI) image in greyscale. 
At the center of PGC 029578 the mean \HI\ velocity is 
$3513 \pm 5$ \kms, which differs considerably from the
long-slit optical value of $3586 \pm 30$ \kms\ for the 
nucleus obtained
by \citet{zaritsky97}. We wonder if the latter is a misprint.
Our \HI\ value of $v_{\rm sys}$ for PGC 029578 is 
$143 \pm 5$ \kms\ below that of
NGC 3145. Thus an encounter between PGC 029578 and
NGC 3145 could also have had a significant component 
perpendicular to the disk of NGC 3145.

Figure\,\ref{PGC.rotcurve} displays the position-velocity (P-V)
diagram of PGC 029578 along its kinematic major axis. 
Out to a radius $R$ of $50''$ (i.e., approximately the 
optical semi-major axis  $D_{25}/2$),
the position angle of the kinematic major 
axis = $270\degr$ on the receding 
side, and the rotation curve is symmetric.
 Beyond $R$ =$50''$, (i) the kinematic major
axis on the receding side appears to align better with the 
isophotal major axis, which has a PA = $283\degr$, and (ii)
the velocity field on the approaching side becomes irregular.

Estimates of $i$ are  $68\degr$ from the optical axis ratio
(see Table\,\ref{basic.optical}) and
$61\degr \pm 2\degr$ from the \HI\ axis ratio. 
At $R$ =$50''$ = 13 kpc, 
$(v - v_{\rm sys}) \sin i$  = 76 \kms. Taking $i$ = 
$64\degr \pm 4\degr$ gives $v_{\rm rot} = 85 \pm 3$ \kms.
Then for a spherically symmetric mass distribution, the
dynamical mass $M_{\rm dyn}$ = $2.2 \times 10^{10}$ \msun.
Out to the same radius ($R =50''$), the \HI\ mass of
PGC 029578 is $2.9 \times 10^9$ \msun, so
$M$(HI)/$M_{\rm dyn}$ = 0.13. 

 From the dynamical masses, the ratio of the mass of 
PGC 029578 to that of NGC 3145 is 0.044. From the $K_s$
luminosities, the ratio of the stellar mass of NGC 3143 to that of 
NGC 3145 is 0.073. Thus the Sdm galaxy PGC 029578 has 
a lower mass than NGC 3143. PGC 029578 is also less luminous
optically than NGC 3143 as it is 1.4 mag fainter in
$B$ and 1.6 mag fainter in $R$ than NGC 3143
(see Table\,\ref{basic.optical}). However, as noted above, 
PGC 029578 has five times as much \HI\ as NGC 3143.
This is an indication that PGC 029578 has not experienced
an encounter recently since otherwise its \HI\ content 
would be lower as a result of gas loss and/or enhanced star formation.

It is more likely that NGC 3143 was involved in a recent interaction 
because its SFR seems to be enhanced,  its \HI\ emission 
is 50\% more extended on its northeastern side,
it is somewhat  
deficient in \HI, and  it appears to have  a large molecular fraction.

 The simple analytic model in the Appendix demonstrates that 
 an encounter between NGC 3143 and NGC 3145 could have triggered the observed
 warping mode in NGC 3145. It illustrates some of the considerations necessary
for making a detailed numerical model. The latter is beyond the scope of
this paper.   The observed anti-symmetric warping motions in NGC 3145
  (see Section 4.2.1)  have line-of-sight
 magnitude $\sim 40$  \kms (or $v_z \sim 60$ \kms\ if the observed motions
are completely perpendicular to the disk). 
In the model, the magnitude $\Delta v_z$ of the warping motions and 
the maximum displacement  $\Delta z$ of the material above the 
plane of the disk depend on
how shallow the attack angle of the companion's trajectory is relative to the disk
of NGC 3145.
For a particular choice of  orbital tilt angle, this analytic model creates warping motions
 in NGC 3145 of magnitude $\sim 200$ \kms, 
  with NGC 3143 pulling up on one side of the disk of NGC 3145
  and down on the opposite side of the disk. These would warp the disk by 
$\sim 5$ kpc. Given the assumptions in the model, these values 
of $\Delta v_z$ and $\Delta z$ are
uncertain by a factor of a few. A slightly steeper angle of attack with the
same flyby velocity would give a more modest warp.  The comparison 
in the Appendix between the time elapsed since closest approach and
rough estimates of the period of vertical oscillation of the warp suggests that
the warp material may have experienced about one vertical oscillation.
Thus the extra-planar features produced by this encounter should still
be present.

\section{Discussion and Conclusions}

NGC 3145 is a spiral galaxy with several optical peculiarities. 
 For example
on the southern side of NGC 3145, stellar arms cross, 
forming ``X''-features, and outline an apparent triangular region. 
 Sandage's peculiar arm forms the eastern edge 
of the {\it triangle}. As this arm heads southwest from the inner disk  of
NGC 3145, it crosses first a string of blue clumps, then
an inner spiral arm at the eastern apex of the {\it triangle}, and then another
stellar arm at the southern apex of the {\it triangle}.

Together with its two smaller companions, NGC 3145
forms the NGC 3145 triplet. To study the above features and other
optical anomalies, we  
analyzed VLA \HI\ observations of this group and VLA \sixcm\ radio
continuum observations of NGC 3145.

Our \sixcm\ radio continuum observations yield the following
information about NGC 3145

(1) Lack of prominent radio continuum emission at the eastern and 
southern apices of the {\it triangle} rules out shock fronts at the
arm-crossing ``X'''s there. This means that the arms  appearing to
cross at these locations must be in different planes and that 
the portion of Sandage's peculiar arm outlining the eastern edge
of the {\it triangle} is not in the disk.

(2) North of the {\it triangle}, there is extended 
radio continuum emission along
Sandage's peculiar arm. This is indicative of shocked gas.
Hence this portion of Sandage's 
peculiar arm, which includes where it crosses the string of blue
clumps, appears to be in the  disk. The suggestion is that
Sandage's peculiar arm emerges from the disk before it
reaches the {\it triangle}.

Our \HI\ observations find the following kinematic 
anomalies in NGC 3145.

(1) In large areas of NGC 3145, the line-profiles are skewed. 
In the middle-to-outer part of the gas disk, line-profiles near
the major axis consist of a large amplitude peak plus a broader, skewed 
wing.   Relative to the disk, the gas in the skewed wing 
has $z$-motions away from us on the approaching side of the 
galaxy and $z$-motions towards us on the receding side. The 
difference $v_2 - v_1$ between the central velocity of the skewed
wing and that of the large amplitude peak has a mean value 
$37 \pm 3$ \kms\ on the approaching side and 
$-47 \pm 7$ \kms\
on the receding side. This kinematic anti-symmetry implies 
that there has been a perturbation with a sizeable component
perpendicular to the disk over 
large spatial scales, e.g., the disk is
undergoing a warping as a result of
a close passage by one of  its companions. We detect these warping motions
out to a radius of at least $105''$ = 28 kpc.

(2) There are two features whose velocities imply that they are
out-of-plane tidal arms. One is the apparent branch of the
main spiral arm on the northeastern side of the galaxy.
Where the {\it branch} and the main spiral arm appear to
intersect in projection, the {\it branch} has a line-of-sight
velocity $\sim 150$ \kms\ greater than that of the
main spiral arm. We conclude that the {\it branch} is an 
out-of-plane tidal arm and may have streaming motions along it.  
The other is
Sandage's peculiar arm as line-profiles of the 
 western-antenna extension of Sandage's peculiar
arm exhibit two clearly distinct peaks of comparable amplitude 
separated by 56 \kms\ in line-of-sight velocity. We attribute one
of these to \HI\ in the disk and the other to \HI\ in Sandage's 
peculiar arm, which thus appears to be another tidal arm.
These arms appear relatively short for tidal arms; more sensitive
\HI\ observations are necessary to see if these tidal arms extend
farther.

(3) The distribution of \HI\ emission from NGC 3145 is not axisymmetric:
within the main optical disk there are massive \HI\ concentrations NW, NE,
and SW of the nucleus and a trough on the SE side. This peculiarity is
other evidence of an encounter.

Our observations solve the puzzle of Sandage's peculiar arm, i.e., it is a
tidal arm tilted with respect to the plane of the disk of NGC 3145, and
this tidal arm emerges from the disk 
before it reaches the apparent arm crossings at the
{\it triangle}.  Furthermore, our \HI\
observations revealed anti-symmetric warping motions in NGC 3145
on large spatial scales out to 28 kpc. 
The two relatively-short, extra-planar tidal arms
 are evidence that NGC 3145 has recently experienced  
a combination of azimuthal and 
perpendicular perturbations. The warping motions are evidence that 
it has recently experienced
a perpendicular perturbation affecting the middle-to-outer  part of the gas disk.

The Sdm galaxy PGC 02978 shows 
no evidence of having undergone an encounter recently, whereas
NGC 3143  has an enhanced SFR, somewhat of an \HI\ deficiency
compared to its SFR,
  \HI\   emission 50\% more extended on its northeastern side
than on the opposite side,
and an apparently large molecular fraction for a small galaxy.
Thus  NGC 3143  is the more likely of the two companions  to
have  interacted with NGC 3145.

Our simple analytic model demonstrates that an encounter between
 NGC 3143 and NGC 3145 is a plausible explanation for the observed 
 warping motions in NGC 3145.

\acknowledgments

We thank the referee for making detailed comments and suggestions that
significantly improved this paper.
The \HI\ and radio continuum images used here are from our observations
in Programs AK 368 and AK 327 at the Very Large Array of Radio 
Telescopes  (VLA). 
The Hubble Space Telescope image used in this research is based 
on observations made with the NASA/ESA Hubble Space telescope and
obtained from the Hubble Legacy Archive, which is a collaboration 
between the Space  Telescope Science Institute (STScI/NASA), the
Space Telescope European Coordinating Facility (ST-ECF/ESA), and the
Canadian Astronomy Data Centre (CADC/NRC/CSA). This publication
makes use of data products from the Widefield Infrared Survey 
Explorer, which is a joint project of the University of California,
Los Angeles, and the Jet Propulsion Laboratory/California Institute
of Technology, funded by the National Aeronautics and Space
Administration.
We also utilized data products from the Two Micron All Sky Survey
(2MASS), which is a joint project of the University of 
Massachusetts and the Infrared Processing and Analysis Center/
California Institute of Technology, funded by NASA and the National 
Science Foundation.
We obtained helpful information from
 the NASA/IPAC Extragalactic Database (NED),
which is operated by the Jet Propulsion Laboratory, California Institute
of Technology, under contract with the National Aeronautics and Space
Administration. 
We thank Paul Eskridge for making observations for us at the 
Michigan-Dartmouth-M.I.T. Observatory. We thank Philip Appleton
for making observations for us at the Fick Observatory of Iowa State
University.


%% in parentheses, after the keyword, but they will not be verified.

\appendix

\section{Analytic plausibility models for the collision}

Although the observations of the NGC 3145 system may not appear sufficient to constrain 
narrowly the parameters of the collision that likely produced the anomalous disk structure
of NGC 3145,  one factor  gives considerable leverage: the companions are far from 
the primary i.e., more than a few primary diameters. It is possible that the 
relative orbit is hyperbolic. Yet a 
second fact, that the mass ratio is evidently less than 10\%, makes that unlikely. A hyperbolic encounter with a low mass companion will not have much effect. A slightly bound orbit is more 
likely. In either case, the effects of dynamical friction are probably minimal. 

We assume that the relative orbit is marginally bound and, based on the arguments
in Section 5, that NGC 3143 is the collision companion. We use its observed properties:
plane-of-sky distance from NGC 3145 = 140 kpc and line-of-sight relative velocity=
125 \kms.
The rotation curve of  NGC 3145 is relatively flat beyond 13 kpc, so we can make the further simplifying assumption that the dark-halo density profile (and logarithmic potential) is the 
same all the way out to the companion. In that case the orbital shape will be well described by 
the p-ellipse approximation of \citet{st06, st15}, 
especially by Figs. 6-8 of the latter paper. Even though we do not know the angular orientations 
of the orbit, knowing its shape is helpful. For example, a very radial p-ellipse in this potential is significantly curved only near its maximum and minimum radii, and is nearly straight otherwise. 
This, in turn, tells us that unless the companion is at its maximum radius (apoapse), its 
direction of motion has changed little since it left the vicinity of the primary. This means 
that the trajectory has not curved or bowed out into the E-W direction, but lies mostly in the plane containing the N-S line, and perpendicular to the sky.

Two interrelated questions remain: 1) what is the angle of the companion's trajectory relative to the plane of the sky and, 2) how close is the companion to apoapse? Given the large visible separation between the two galaxies, and the fact that particles on very radial orbits spend most of their time near apoapse, it seems reasonable to suppose that the companion is indeed near apoapse. The following energy equation provides a constraint,

% eq A1
\begin{equation}
\frac{1}{2} {v_1}^2 = -\frac{GM_1}{r_1}
ln \left[ \frac{r_1}{r_{\rm max}} \right],
\label{eqA1}
\end{equation}
where $v_1$ is the current relative velocity of the companion, so the left-hand side of the equation is the specific kinetic energy. The right-hand side is the potential energy difference between the current radius $r_1$ and the apoapse radius $r_{\rm max}$.  Here $M_1$ is the mass contained within the radius $r_1$. Specifically, 

% eq A2
\begin{equation}
M_1 = M_o \left( \frac{r_1}{r_o} \right),
\label{eqA2}
\end{equation}
where the `o' subscript refers to values at the radius of NGC 3145
beyond which the trajectory of the companion is assumed to be nearly straight until it reaches apoapse.  From Table\,\ref{HIproperties} we take $r_o$ = 33 kpc and 
$M_o$ = $5 \times 10^{11}$ \msun.

Equation A1
can be solved for the radius ratio, and with equation A2 we have,

% eq A3
\begin{equation}
ln \left [ \frac{r_1}{r_{\rm max}} \right] = \frac{-r_1 {v_1}^2}{2GM_1}  = \\
 -0.12 \left( \frac{r_o}{33\ {\rm kpc}} \right)
    \left( \frac{v_1}{125\ {\rm km\  s^{-1}}} \right)^2
%\times
\left( \frac{5 \times 10^{11}\ M_{\odot}}{M_o} \right)
(\sin \theta)^{-2} ,
\label{eqA3}
\end{equation}

\noindent where the sine term accounts for the angle between the trajectory and the plane of the sky, and the fact that we observe the velocity component perpendicular to it. 
For example, if we choose $\theta = 45^{\circ}$, 
then $r_1  = (140\ {\rm kpc}) / \cos(45\degr)$ = 200 kpc,
$v_1 = (125\ {\rm km\ s^{-1}}) / \sin(45\degr)$ = 177 \kms, 
and the radius ratio is $0.79$. This value for the radius ratio is consistent 
with the assumption that the companion is near apoapse. 

Then we can use the energy equation again to estimate the velocity at $r_o$, which is near closest approach. i.e., 

% eq A4
\begin{equation}
    {v_o}^2 = \left( 177\ {\rm km\ s^{-1}} \right)^2
 + \frac{2GM_o}{r_o} ln \left[ \frac{r_1}{r_o} \right].
\label{eqA4}
\end{equation}

 This example calculation gives $v_o$ = 515 \kms, which seems reasonable.

We do not know the angle of attack of the companion trajectory relative to the NGC 3145 disk. However, given that the disk does not have a strong ring, the impact was unlikely to be
 perpendicular.  NGC 3145  has  two features that are extra-planar tidal arms, one in the
northeast and the other in the south to southwest, and neither is a prominent, long tidal tail.
Therefore the impact was unlikely to be planar.  The combination of the two stubby, out-of-plane
 tidal arms (evidence for azimuthal and  perpendicular perturbations) and the 
global warp suggests an
intermediate angle of attack.

With estimates for the maximum radius of about 250 kpc, and the minimum radius of about 
15 kpc, we can use the p-ellipse approximate equation to estimate the p-ellipse eccentricity in this potential, 

% eq A5
\begin{equation}
\frac{r_{max}}{r_{min}} = 
\left( \frac{1+e}{1-e} \right)^{1/2}.
\label{eqA5}
\end{equation}

This yields $e =0.993$. The formulae of \citet{st06, st15}, suggest that such an orbit will turn 
through an angle of about $125^{\circ}$ near closest approach. Thus, we can envision its closest 
approach arc as roughly semi-circular ($\pm 60^{\circ}$ on either side of the impact point), with a radius of about $15$ kpc  and a fairly shallow angle of impact. 
At the ends of this arc the companion is about one radius length above or below
(different sectors of) the disk. 
On the incoming end it will pull up. On the outgoing end it will pull down. 

Assume the companion pulls on each of these disk sectors for a time roughly equal to 
the time it takes to traverse half the curved arc, i.e., a time about equal to about
(15 kpc/470 \kms) $\simeq 32$ Myr. Then, the vertical velocity impulse felt by that sector is 
about,

% eq A6
\begin{equation}
\Delta v_z = a_{comp} \Delta t \simeq
\frac{GM_{comp} \Delta t}{\left( \frac{r_{\rm min}}{3} \right)^2} \simeq
200\ {\rm km\ s^{-1}},
\label{eqA6}
\end{equation}

\noindent where the companion mass has been assumed to be about 
$3.7 \times 10^{10} M_{\odot}$, and it is assumed that the mean vertical acceleration 
occurs when the companion is at a z distance equal to about $1/3$ of $r_{min}$, i.e., a moderately shallow angle of attack relative to the disk plane. 

Assuming that this impulse is countered by the vertical component of the halo potential, material in this sector will rise to a height of about, 

% eq A7
\begin{equation}
\Delta z \simeq \frac{\Delta v_z^2}
{ \left( \frac{2GM(r_{min})} {r_{min}^2} \right)}
 \simeq 5.0\ {\rm kpc.}
\label{eqA7}
\end{equation}

This shows that it is plausible that the encounter has warped the disk of NGC 3145 by
 at least several kpc.
At the given flyby velocity, a slightly steeper angle than that assumed in
eq. (A6) would give a more modest warp,
 while a shallower angle  would give an even stronger warp.
If the flyby velocity was much greater, then the material would not be lifted to kpc levels. 
If the flyby velocity was much lower, with the moderate angle of attack, longer tails would be produced. 

There are also some timescale constraints, for example, how does the time since closest 
encounter compare to the oscillation time of the warp? If the latter is much shorter 
than the former, we might expect the warp to have smoothed away by the present. 
As an estimate for the former we can use the half period of the adopted p-ellipse orbit. 
This period is given by eq. (C4) in Appendix C of  \citet{st06} (with a small error
corrected).
In the present notation, the half-period can be written as, 

% eq A8
\begin{equation}
T_{1/2} = 1.1 \left( \frac{5 \times 10^{11}\ M_{\odot}}{M_o} \right)
\left( \frac{r_o}{33\ {\rm kpc}} \right)
\left( \frac{a}{132\ {\rm kpc}} \right)^2\ {\rm Gyr}
\label{eqA8}
\end{equation}

\noindent where $a$ is the semi-major axis of the orbit ($(15 + 250)/2 = 132$ kpc). 
This value is also the same as the free-fall time onto the galaxy center with the
 assumed parameters.

The oscillation time can be estimated by the vertical epicyclic frequency of the disk. 
Since most of the
warp is relatively far out in the disk,  we assume that the local gravity of the disk 
there is negligible and the restoring force is provided by the vertical component of the 
total gravity interior to that radius. This approximation seems especially appropriate 
for a large warp involving most of the local disk material. Then we can approximate the 
vertical restoring gravitational acceleration as, 

% eq A9
\begin{equation}
g_z = -\frac{\partial \Phi}{\partial z}
\simeq - \frac{GM(r)}{r^2} \frac{z}{r}.
\label{eqA9}
\end{equation}

\noindent The vertical epicyclic frequency is defined by, 

% eq A10
\begin{equation}
\nu^2 = \frac{\partial^2 \Phi}{ \partial z^2}
\simeq \frac{GM(r)}{r^2 + z^2} 
\simeq \frac{GM(r)}{r^2}.
\label{eqA10}
\end{equation}

\noindent Assuming a nearly flat rotation curve, we approximate $M(r = 15\ {\rm kpc})$ as 
about half $M_o$, and obtain the following estimate of the vertical epicyclic period, 

% eq A11
\begin{equation}
\tau = \frac{2 \pi}{\nu}
\simeq 352\ {\rm Myr}. 
\label{eqA11}
\end{equation}

\noindent Taken at face value the comparison of eqs. (A8) and (A11) suggest that the warp has 
gone through a few epicyclic oscillations by the present. This is possible, but it is 
more likely that the oscillation period is considerably longer. The first reason for this is 
that the epicyclic approximation assumes a very small excursion. Since the estimate of 
eq. (A7) yields a fairly large excursion, the true nonlinear oscillation period is likely 
much longer. The second reason is that once the  material in the warp moves out 
of the disk plane it will not be centripetally balanced in the radial direction. Its true motion 
(even in an azimuthally co-rotating frame) will be a combination of nonlinear vertical and 
radial oscillations. This will introduce another significant increase in the overall period. 
Thus, it is likely that the vertical epicyclic timescale should be increased by a factor of a 
few, and the warp material has experienced about one oscillation. 

These timescale considerations further constrain the putative encounter, and 
especially the angle-of-attack. In sum, if NGC 3143 is indeed the collision companion, then the structure of the encounter is quite constrained.

\end{document}